\documentclass[runningheads]{llncs}
\usepackage{booktabs}  
\usepackage{multirow}  
\usepackage{adjustbox}
\usepackage{siunitx}
\usepackage{graphicx}
\usepackage{caption}
\usepackage{subcaption}
\usepackage{xurl}
\usepackage[hidelinks]{hyperref}
\usepackage{amsmath}
\usepackage{color}
\usepackage{balance}
\usepackage[misc]{ifsym}
\usepackage{orcidlink}
\usepackage[numbers,sort]{natbib}
\usepackage{subcaption}
\usepackage{makecell}
\usepackage{pifont}

\def\BibTeX{{\rm B\kern-.05em{\sc i\kern-.025em b}\kern-.08em
    T\kern-.1667em\lower.7ex\hbox{E}\kern-.125emX}}

\begin{document}
%

\title{Lightweight Detection of Electromagnetic Signal Injection Attacks on Image Sensors}

\titlerunning{Lightweight Detection of ESIA on Image Sensors}

\author{Youqian Zhang\inst{1}\orcidlink{0000-0003-0907-7998}\textsuperscript{\Letter} \and
Chunxi Yang \inst{2}\orcidlink{0009-0001-5435-6083} \and
Eugene Yujun Fu \inst{2}\orcidlink{0000-0003-1048-1904} \and
Sze Yiu Chau \inst{3}\orcidlink{0000-0001-9300-0808} \and
Haibo Hu \inst{1}\orcidlink{0000-0002-9008-2112} \and
Xiapu Luo \inst{1}\orcidlink{0000-0002-9082-3208}}

\authorrunning{Y. Zhang et al.}
%
\institute{The Hong Kong Polytechnic University \and
The Education University of Hong Kong \and
Simon Fraser University \\
\textsuperscript{\Letter}\email{you-qian.zhang@polyu.edu.hk}
}
\maketitle
\begin{abstract}
Electromagnetic signal injection attacks (ESIA) pose a growing threat to image sensors, which are increasingly used in different intelligent systems. 
By emitting electromagnetic interference, adversaries can manipulate pixel values, potentially misleading downstream artificial intelligence (AI) models and causing unsafe decisions in these systems. 
We present a lightweight detection method that leverages optically black pixels, which are non-exposed pixels already present in many modern image sensors, to identify the attacks. 
Our detection approach achieves an area under the receiver operating characteristic curve (ROC-AUC) of up to 99.6\% and an Equal Error Rate (EER) as low as 0.027 across diverse attack conditions. 
Our method requires minimal computational overhead and no hardware modifications, making it a practical and effective defense for securing vision-based systems against ESIA.
\end{abstract}


\section{Introduction}
\label{sec:introduction}

Image sensors are important in modern intelligent systems, enabling computer vision applications in autonomous driving~\cite{ap2017baidu,waymo2025,cruise2025}, intelligence in robotics~\cite{shadowhand2025,paxini2025}, and surveillance~\cite{adt2025,ring2025}. 
These systems rely on visual data to make critical decisions, making the integrity of image sensor outputs essential for safety and security. 
However, recent research has revealed that image sensors are vulnerable to \textit{electromagnetic signal injection attacks (ESIA)}, where adversaries emit electromagnetic interference to manipulate pixel values~\cite{jiang23glitchhiker,zhang2024esia,zhang2024modeling,liao2025your,kohler2022signal,liu2025magshadow,ren2025ghostshot}.
In this work, we focus on Complementary Metal-Oxide-Semiconductor (CMOS) sensors due to their much more widespread use across various applications than other sensing technologies such as Charge-Coupled Device (CCD)~\cite{insights2022cmos}. 
Unless stated otherwise, the term ``image sensor'' hereafter refers to a CMOS image sensor.

\begin{figure}[t]
    \centering
    \begin{subfigure}{0.32\textwidth}
        \centering
        \includegraphics[width=\textwidth]{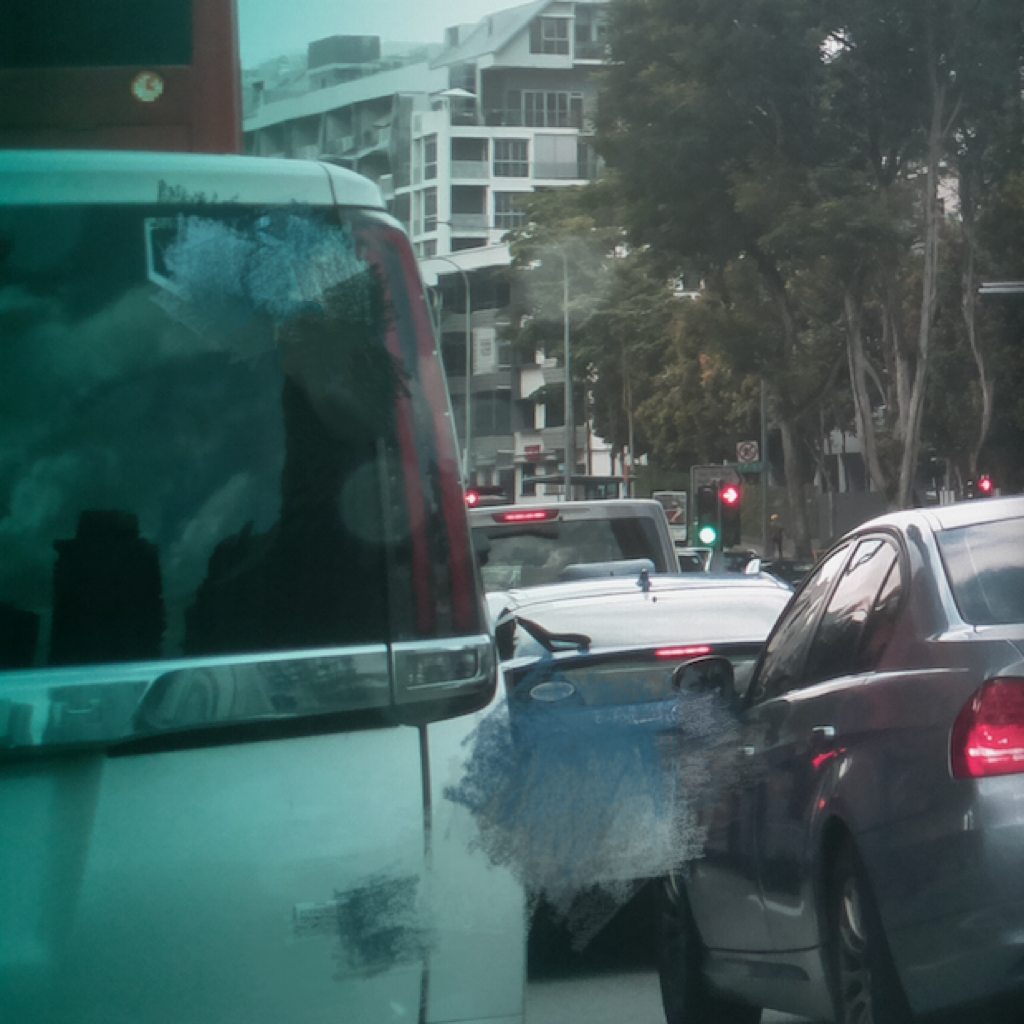}
        \caption{Without Attack}
    \end{subfigure}
    \begin{subfigure}{0.32\textwidth}
        \centering
        \includegraphics[width=\textwidth]{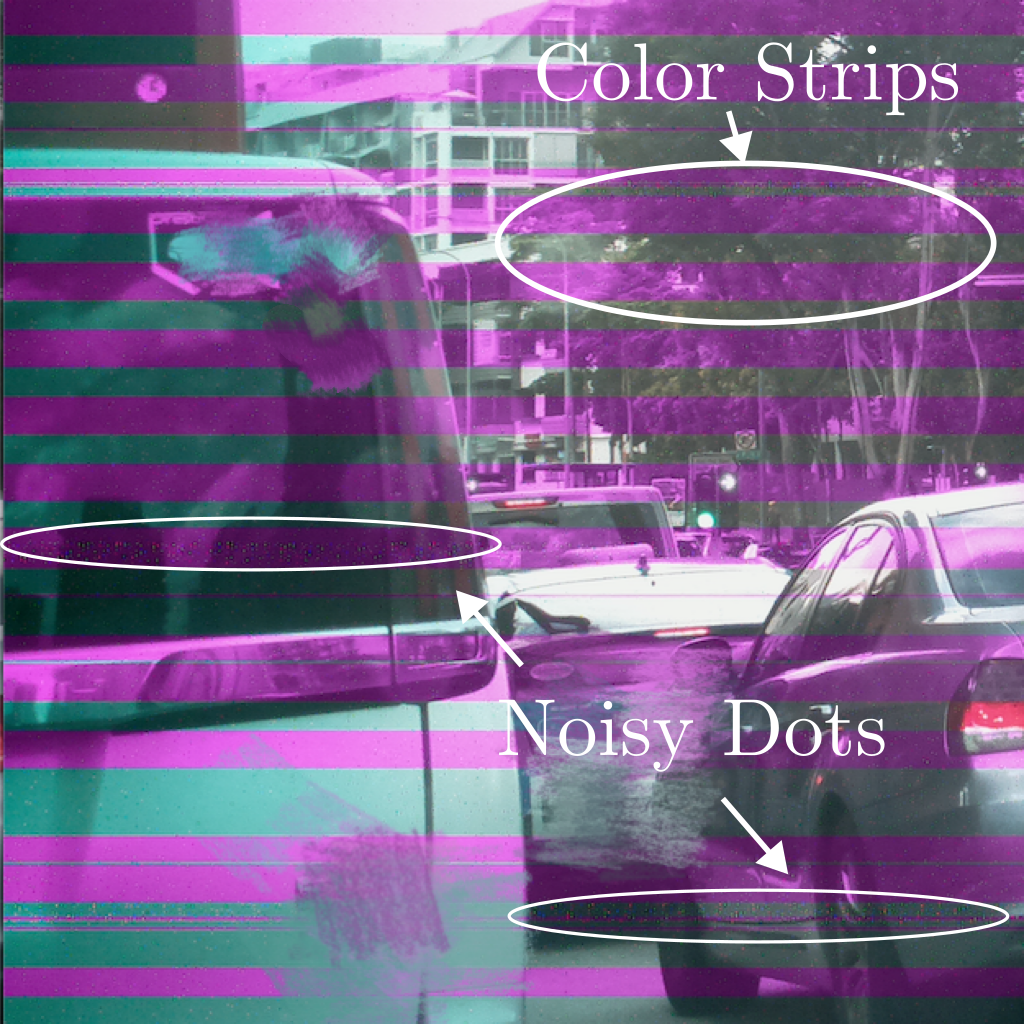}
        \caption{With Attack}    
    \end{subfigure}
    \caption{The effect of electromagnetic signal injection on active pixel regions: (a) the normal frame; (b) the attacked frame with visible color strips and noise.}
    \label{fig:street_wo_w_atk}
\vspace{-2em}
\end{figure}

Unlike other physical attacks, such as adversarial stickers or patches on target objects~\cite{guesmi2023physical}, directing lasers~\cite{petit2015remote,yan2016can,fu2021remote,wang2021can,kohler2021they,yan2022rolling,man2024remote}, or ultrasound beams at image sensors~\cite{ji2021poltergeist,cheng2023adversarial,zhu2023tpatch}, electromagnetic signal injection is more flexible.
Specifically, it does not require physical manipulation of the target object, operates without direct line-of-sight since electromagnetic radiation can penetrate physical barriers (similar to Wi-Fi), and can simultaneously disrupt multiple targets. 
Such attacks interfere with the image sensor’s readout process, leading to two primary types of disruptions, as illustrated in Figure~\ref{fig:street_wo_w_atk}: (1) the corruption of pixel values, resulting in noisy dots within the image, and (2) the loss of entire rows during transmission, leading to color strips caused by incorrect color interpolation during image reconstruction. 
Such disruptions can alter or obscure visual content in the captured frames, potentially misleading downstream computer vision algorithms, including object detection, segmentation, depth estimation, face recognition, and image classification~\cite{jiang23glitchhiker,zhang2024esia,zhang2024modeling,liao2025your}. 
Such attacks have caused task performance to degrade substantially, in some cases by over 50\%~\cite{zhang2024modeling}. 
In safety-critical domains like autonomous driving, these manipulations may result in dangerous consequences, such as misinterpretation of traffic scenes, and reduction of drivable areas~\cite{liao2025your}.

Given that the impacts of ESIA on image sensors have been extensively studied and empirically validated in prior work, this paper does not aim to reanalyze those effects. 
However, there are currently no effective defense mechanisms, leaving many image-sensor-based systems vulnerable to such attacks.
Note that while electromagnetic compatibility (EMC) standards~\cite{paul2022introduction} are a core part of protecting electronic systems from unintentional interference, they are not sufficient to defend against deliberate ESIA. 
Prior research mentioned above has shown that even systems compliant with EMC regulations, i.e., equipped with shielding, filtering, and grounding, remain vulnerable. 
This highlights the need to go beyond EMC compliance and develop active defense mechanisms specifically tailored to such threats.

In order to defend against ESIA, detection plays a critical role in enabling systems to recognize that an attack is currently taking place and respond accordingly. Our work here thus focuses on devising a practical and effective detection mechanism. This, however, is far from trivial, due to several fundamental challenges listed below.

\textbf{C1: Lack of Lightweight Detection Methods.}
While a few detection strategies have been proposed, such as using external sensors~\cite{tu2021transduction}, these approaches are not designed for image sensors and often come with hardware changes, implying more deployment costs (see detailed discussion in Section~\ref{sec:related_work}). 
As a result, there is currently no reliable and lightweight detection solution that can be easily integrated into image sensor systems without additional hardware.

\textbf{C2: Absence of Ground Truth in Active Pixels.}
Although the attacks directly affect the values of active pixels (i.e., the pixels used to form the actual image), these values are inherently scene-dependent, as they fluctuate due to changing lighting conditions, object textures, motion, and other environmental factors. 
This makes it challenging to determine whether a given pixel value has been manipulated or is simply a result of normal variation (see detailed discussion in Section~\ref{sec:comparison_with_ml_and_dl_classifers}).

\textbf{C3: Diverse Sensors and Environmental Conditions.}
Image sensors differ widely in terms of their architecture, resolution, and noise characteristics. 
Moreover, the impact of electromagnetic interference is sensitive to system-specific and environmental factors, such as system configurations and attack parameters.
This diversity creates a significant challenge: a reliable detection method should work consistently across different sensor models and attack conditions. 
Designing a method that maintains high detection performance under such varied conditions is non-trivial and has not been addressed effectively in prior work.

To address the above challenges, our work makes the following key contributions:

\textbf{S1: System and Threat Modeling.} We begin by abstracting a high-level system model that captures the common architecture of intelligent systems using image sensors for perception. 
  We also define a realistic adversary model that outlines the capabilities and limitations of the attackers. (Section~\ref{sec:system_model_and_adversary_model})
  
\textbf{S2: Novel Detection Approach Using Optically Black Pixels.} We propose a lightweight detection method that leverages \textit{optically black pixels}, which are originally designed for black level calibration, as a built-in monitoring channel. By analyzing statistical variations in these non-illuminated pixels, the system can reliably detect the presence of signal injection attacks with a provable security guarantee. (Section~\ref{sec:detection_method})
  
\textbf{S3: Validation Across Different Sensors and Conditions.} We implement and evaluate our method in real-world hardware setups under a wide range of attack conditions. Our results demonstrate that the proposed detection strategy is both effective and robust, achieving high accuracy across diverse sensor models and attack scenarios. (Section~\ref{sec:experiements} and Section~\ref{sec:analysis_of_detection_performance}).

The rest of this paper is organized as follows.
Section~\ref{sec:background} provides background on electromagnetic signal injection attacks.
Section~\ref{sec:discussion} discusses broader implications of the proposed approach.
Section~\ref{sec:related_work} reviews related work, and a conclusion will be drawn in Section~\ref{sec:conclusion}.
A repository of our code and non-sensitive dataset is available at \url{https://osf.io/kfv9h/overview?view_only=30799e9f9cd34e14aabf408686871e12}.

\section{Background}
\label{sec:background}

Electromagnetic signal injection attacks (ESIA) are an emerging and critical threat to a wide range of electronic systems. 
Recent studies have demonstrated their feasibility and impacts on various devices/systems, including smartphones (e.g.,~\cite{wang2022ghosttouch,maruyama2019tap,gao2023practical}), smart speakers (e.g.,~\cite{xu2021inaudible,fokkens2021machine}), drones (e.g.,~\cite{kim2022review,jang2023paralyzing,dayanikli2022physical}), implantable medical devices (e.g.,~\cite{kune2013ghost,rasmussen2009proximity}), and even critical infrastructure, such as power systems (e.g.,~\cite{yang2024rethink,barua2020hall,szakaly2023assault}), etc. 
The increasing dependence on these systems for daily activities and critical services makes the attacks a growing concern, as they can disrupt, manipulate, or compromise target systems with potentially severe consequences.

At the core of the attacks is an exploitation of unintended antenna-like behavior in electronic circuits. 
Metal traces and wires in printed circuit boards (PCBs) can unintentionally receive electromagnetic energy from their environment~\cite{wilson2010radiation,paul2022introduction}. 
When an attacker transmits a carefully crafted electromagnetic signal, these conductive paths may pick up the energy, resulting in induced voltages or currents that interfere with legitimate operations. This allows attackers to influence system behavior without physical contact or software compromise.

Previous research mentioned above has demonstrated that these attacks can be executed using readily available, commercial off-the-shelf (COTS) equipment such as signal generators, amplifiers, and antennas, meaning that attackers do not require specialized knowledge or highly sophisticated tools to carry out an attack. 
To achieve a successful attack, an attacker essentially needs to tune the frequency and the power of the electromagnetic signal~\cite{yan2020sok}.
The attacker typically tunes the frequency to match the resonant frequency of the injection point, such as the wires/traces~\cite{kune2013ghost}.
At the resonant frequency, the efficiency of power transfer from the electromagnetic wave to the circuit is maximized, allowing the injected signal to induce significant interference with minimal effort.
Higher power signals are more likely to induce more significant effects.
However, attackers must strike a balance: excessive power risks damaging the circuit (e.g., frying components), which could make the system inoperable.

\section{System Model and Adversary Model}
\label{sec:system_model_and_adversary_model}

In this section, we present a high-level system model that captures the key components and signal flow of typical intelligent systems employing image sensors. 
We then present the adversary model, describing attackers' capabilities and constraints. 

\subsection{System Model}
\label{sec:system_model}

\begin{figure*}[t]
\centering
\includegraphics[width=1\textwidth]{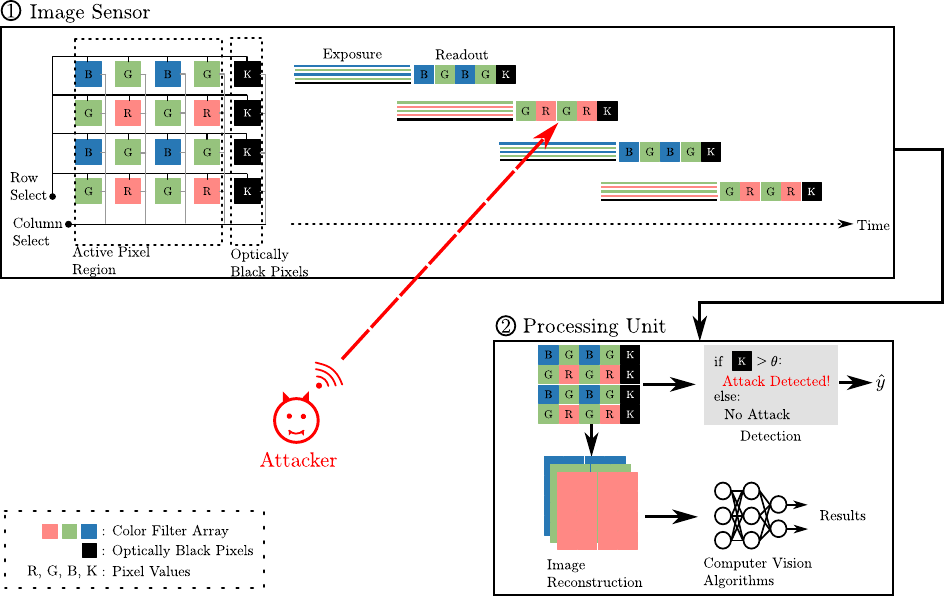}
\caption{The system consists of an image sensor and a processing unit, which can capture, convert, and process visual information.}
\label{fig:system_model}
\vspace{-2em}
\end{figure*}

The system consists of two primary components: an \ding{172} Image Sensor and a \ding{173} Processing Unit, as depicted in Figure~\ref{fig:system_model}. 
These components are connected by metal conductors (e.g., electrical cables or PCB traces), and they work together to capture, convert, and process visual information, enabling advanced analysis and interpretation of the captured scene (see Figure~\ref{fig:system_model}). 

\subsubsection{Image Sensor.}
The image sensor captures visual information from the environment and converts it into digital signals. 
It is composed of (millions of) photodiodes, each acting as a light-sensitive element arranged in a grid-like structure. 
When exposed to light, a photodiode accumulates an electrical charge proportional to the intensity of the incident light, forming the brightness value of the pixel. 
The photodiodes are divided into two key regions: the active pixel region, which is responsible for capturing the image, and the optically black pixels, which are entirely covered to prevent light exposure. 
It is important to note that active pixels and optically black pixels are identical, with the only difference being that optically black pixels are never exposed to light.

Regarding the active pixel region, each photodiode corresponds to a single pixel in the image. 
To support color imaging, a Color Filter Array (CFA) is placed over the photodiode array. 
The CFA filters incoming light so that each photodiode is sensitive to only one specific color component. 
For example, a common CFA used in image sensors is the Bayer filter~\cite{bayer1976color}, as shown in Figure~\ref{fig:system_model}, which follows a repeating 2x2 pattern of one red (R) filter, one blue (B) filter, and two green (G) filters.

The optically black pixels (denoted as $K$ in Figure~\ref{fig:system_model}) are used for black level calibration. 
Ideally, they should output zero in complete darkness. However, due to noise such as thermal noise (e.g., dark current) in the analog front-end, these pixels exhibit non-zero baseline values~\cite{carrere2014cmos,el2005cmos}. 
These offsets are compensated during image reconstruction later to ensure accurate rendering.

The operation of the image sensor consists of two main phases: exposure and readout. 
During exposure, the active pixel region is exposed to incident light for a specific duration, during which each photodiode accumulates an electrical charge based on the intensity of light it receives.
In parallel, the optically black pixels undergo the same exposure duration, but in darkness.
During readout, the accumulated charge from each photodiode is transformed into a digital value via analog-to-digital converters (ADCs), then transmitted as a stream of raw pixel data to the processing unit.

Different types of image sensors employ various techniques for exposure and readout. 
In this work, we consider a common technique that uses rolling shutter exposure and overlapping readout~\cite{basler2025overlapping}. 
As shown in Figure~\ref{fig:system_model}, rolling shutter exposure captures rows of photodiodes sequentially, exposing one row at a time. 
Overlapping readout allows exposure and readout to occur simultaneously across adjacent rows of photodiodes. 
For example, one row is exposed while the previous row’s data is read out, and such an overlapping mechanism can improve frame rate performance.

\subsubsection{Processing Unit.}

The processing unit receives raw sensor data and performs a sequence of operations to reconstruct and analyze the image. 
The first step is black-level calibration, where the output from optically black pixels is used to correct the baseline offset in active pixels.
Next, to reconstruct the full-color image, a demosaicing algorithm is applied. 
This algorithm interpolates the missing color information for each pixel by analyzing the color values of neighboring pixels, based on the CFA’s arrangement (e.g., the Bayer pattern). 
This process, named ``Image Reconstruction'' in Figure~\ref{fig:system_model}, restores the full RGB (Red, Green, Blue) color information for each pixel, producing a complete image.
Following reconstruction, the image may undergo enhancement operations such as noise reduction or contrast adjustment. 
In intelligent systems, the image is typically passed to some Computer Vision algorithms for further analysis, which may include object detection, classification, or semantic segmentation. 
The results of this analysis are then used to inform downstream decisions or control actions, such as navigation or actuation.

\subsection{Adversary Model}

We consider an adversary whose goal is to distort the image data captured by a target system, thereby influencing the output of downstream computer vision algorithms. 

\subsubsection{Attack Mechanism.}  
The attacker exploits the antenna-like behavior of metal conductors (e.g., PCB traces or cables) that connect the image sensor and the processing unit. 
By transmitting electromagnetic waves, the adversary induces malicious voltages on these conductors, thereby corrupting the transmitted pixel data. 
These corrupted values can lead to visible artifacts in the final image, including noisy pixels and color strips (as introduced in Section~\ref{sec:introduction} and Figure~\ref{fig:street_wo_w_atk}).
 
The attacker has access to commercially available radio-frequency (RF) equipment, including signal generators, amplifiers, and antennas. 
They have detailed knowledge of the data protocol and can generate arbitrary analog waveforms, allowing manipulation of the image content.
However, the attacker does not have physical access to the internal components of the system and cannot modify hardware, firmware, or software. 
All attacks are conducted externally using electromagnetic emissions, where the attack remains non-invasive.

\subsubsection{Attack Parameters.}  
To manipulate the sensor readout more precisely, the attacker needs to carefully tune several physical-layer parameters. These parameters govern how effectively the electromagnetic signal couples into the victim system, and they also shape the visual artifacts that appear in the final image. 
We characterize each attack instance using the following physical-layer parameters:

\begin{itemize}
    \item Frequency: The frequency of the attack signal, typically tuned to resonate with circuit elements to maximize coupling efficiency.
    \item Power: The transmission power of the signal, which affects the strength of the induced interference.
    \item Duration: The length of time the attacker transmits the signal.
    \item Distance: The physical separation between the attacker's antenna and the target system, which influences signal attenuation.
    \item Antenna Angle: The orientation of the transmitting antenna relative to the target, which affects the directionality and strength of signal coupling.
\end{itemize}

These parameters are grounded in prior work~\cite{jiang23glitchhiker,yan2020sok}, and we will explore their impacts on our detection method in Sections~\ref{sec:experiements} and~\ref{sec:analysis_of_detection_performance}.

\section{Detection Method}
\label{sec:detection_method}

Building upon the system and adversary models presented in the previous section, we now introduce our proposed detection method for identifying ESIA targeting image sensors.

\subsection{Detection Using Optically Black Pixels}

While reproducing ESIA in our experiments (details in Section~\ref{sec:experimental_setup}), we observed an intriguing phenomenon: in areas of the image that were supposed to be dark, unexpected bright or colorful spots would appear. 
These anomalies were caused by the electromagnetic waves injected by the attacker, which altered the pixel values from what they should have been. 
This observation provided a crucial hint: by monitoring dark regions of an image for abnormal pixel values, it might be possible to detect the presence of such attacks. 
However, in practice, not all images contain sufficiently dark regions, and relying on this as the sole detection method would be unreliable due to the dynamic nature of scenes.

To address this limitation, we turn to optically black pixels. 
Recall that, unlike the active pixels that respond to external light, optically black pixels are shielded from light exposure and thus remain unaffected by changes in external lighting. 
Their values are meant to remain stable under benign conditions, as they are not influenced by the scene being captured. 
This makes them an ideal channel for detecting abnormal variations caused by ESIA. 

Recall the principle of image sensor exposure: all photodiodes in a row, including those in the active region and the optically black pixels, are exposed simultaneously. 
\textit{This means that any ESIA would affect both regions at the same time.}
For example, as illustrated in Figure~\ref{fig:system_model}, an attacker aims to manipulate the second packet (e.g., the GRGRK). 
However, due to the temporal overlap in sensor operation, the exposure of the third row has already begun. 
Consequently, the injected signal influences the optically black pixels in both the second and the third row.
If the values of the optically black pixels deviate from their expected stable baseline, it is a strong indicator that an attack is taking place.

\subsection{Modeling Detection}
In the absence of an attack, the optically black pixel is determined only by noise:
\(
K = n,
\)
where \( n \) is the noise. 
The noise \( n \) is assumed to follow a probability distribution with a cumulative distribution function (CDF) \( N(x) \), defined as:
\begin{equation}
N(x) = \Pr[n \leq x] = \int_{0}^{x} p(u) \, du,
\end{equation}
where \( p(u) \) is the probability density function (PDF) of the noise. For \( 0 \leq \epsilon < 1 \), we can find \( x \) such that
 \(   
 N(x) = \epsilon.
 \)
The smallest such \( x \) is given by:
\begin{equation}
N^{-1}(\epsilon) = \inf \{x \geq 0 : N(x) = \epsilon\}.
\end{equation}
For a chosen (small) tolerance $\epsilon$, a detection threshold \( \theta \) is selected such that:
\begin{equation}
   \theta = N^{-1}(\epsilon). 
\end{equation}
The detection principle is straightforward, as depicted in Figure~\ref{fig:system_model}. 
The detection result $\hat{y} \in \{0,1\}$ is determined by comparing the optically black pixel value $K$ against the detection threshold $\theta$:
\begin{equation}
\hat{y} =
\begin{cases}
1, & \text{if } K > \theta \quad \text{(attack detected)} \\
0, & \text{if } K \leq \theta \quad \text{(no attack)}
\end{cases}
\end{equation}

\subsection{Probability of Bypassing Detection}
\label{sec:detection_method_probability_of_bypassing_detection}

Suppose an attacker injects a signal that raises the black pixel's value by \(\delta\), which depends on both the attack signal characteristics and the physical coupling to the sensor. 
After the injection, the new pixel value becomes
\(
    K = n + \delta.
\)
To remain undetected under our threshold-based mechanism, the attacker’s modified pixel value must not exceed \(\theta\). Thus, the attacker must satisfy:
\(
    K \;\leq\; \theta
    \Longrightarrow
    n + \delta \;\le\; \theta
    \Longrightarrow
    n \;\le\; \theta - \delta.
\)
From the definition of the cumulative distribution function (CDF) \(N(x)\) of the noise \(n\), the probability that this condition holds for a \emph{single} row of optically black pixel is:
\begin{equation}
    \Pr[K \le \theta] 
    \;=\; \Pr[n \le \theta - \delta] 
    \;=\; N\bigl(\theta - \delta\bigr).
\end{equation}

Practical image manipulation often requires an attacker to alter multiple rows of pixels, and let this number be \(r\). 
Assuming approximate independence between rows of pixels, the probability of evading detection across \emph{all} \(r\) manipulated rows is:
\begin{equation}
    \bigl(N(\theta - \delta)\bigr)^{r}.
\end{equation}
Since \(N(\theta - \delta) < 1\) for any \(\delta > 0\), the above term decays exponentially as \(r\) increases. Therefore, the attacker’s probability of bypassing detection becomes negligibly small as the detection involves more rows.
\section{Attack Impacts on Optically Black Pixels}
\label{sec:experiements}

This section evaluates the impacts of attacks on optically black pixels in different conditions.

\subsection{Experimental Setup}
\label{sec:experimental_setup}

\begin{figure}[t]
\centering
\includegraphics[width=0.5\textwidth]{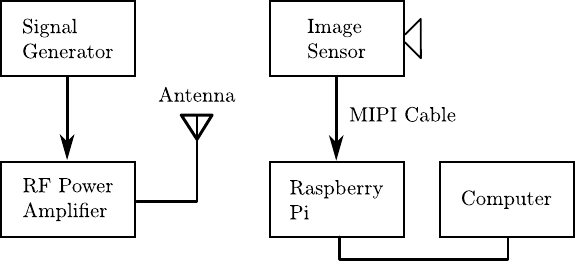}
\caption{An experimental setup for data collection.}
\label{fig:experiment_setup}
\vspace{-2em}
\end{figure}

We constructed an experimental setup, as illustrated in Figure~\ref{fig:experiment_setup}.
For the system component, we used a Raspberry Pi equipped with a compatible image sensor (by default OV5647, but others will be tested as well), connected via an MIPI cable. 
The Raspberry Pi collects the captured raw images, and reconstructs them into full color images.
A computer is used to process and analyze the images.
For the attack equipment, we utilized a signal generator to produce attack signals, and we selected a sinusoidal waveform, which is commonly used in other ESIA that are discussed in Section~\ref{sec:background}. 
These signals were amplified using a radio-frequency (RF) amplifier and transmitted via an omnidirectional antenna, which radiates equal radio power in all directions. 

\begin{table}[t]
\centering
\caption{Summary of Experimental Parameters}
\label{tab:experiment-params}
\resizebox{0.6\columnwidth}{!}{%
\begin{tabular}{ll}
\toprule
\textbf{Parameter} & \textbf{Values / Range} \\
\midrule
\textbf{Frequency} & \SIrange{25}{100}{\mega\hertz} (in \SI{5}{\mega\hertz} steps) \\
\textbf{Power} & \SIrange{1}{3.5}{\watt} (in \SI{0.5}{\watt} steps), and \SI{5}{\watt} \\
\textbf{Duration} & \SIlist{6.7;13.3;20;27;33.3}{\milli\second} \\
\textbf{Distance} & \SIlist{0.05;0.10;0.15;0.20}{\meter} \\
\textbf{Antenna Angle} & \SIlist{0;30;60;90}{\degree} \\
\textbf{Cable Length} & \SIlist{16;30;50;75;100}{\centi\meter} \\
\textbf{Sensor Temp.} & \SIlist{14;30;40;50}{\celsius} \\
\textbf{Sensor Models} & OV5647, IMX219, IMX378, IMX708 \\
\bottomrule
\end{tabular}
}
\vspace{-1em}
\end{table}

In our setup, we successfully induced visible effects on the captured images (as shown in Figure~\ref{fig:street_wo_w_atk}) by injecting the attack signal into the cable connecting the image sensor and the Raspberry Pi, demonstrating the effectiveness of the setup.
Note that our experiments were conducted in a laboratory; when no attack is present, no malicious effect is observed.

There is neither an oracle nor an automated mechanism to precisely determine the minimal configurations of an attack that causes a minor disturbance to images without triggering detection.
Therefore, to evaluate the impacts of the attacks on optically black pixels, we systematically varied key physical and environmental parameters that influence electromagnetic signal injection.
A summary of the parameters is shown in Table~\ref{tab:experiment-params}, which were selected within the capabilities of our equipment.
Unless stated otherwise, experiments are conducted using a default configuration with the following attack parameters: the attack frequency is swept across the specified range; the transmission power is set to \SI{3}{\watt}; the signal duration is \SI{33.3}{\milli\second}; the distance between the transmitting antenna and the image sensor is \SI{0.05}{\meter}; and the antenna is oriented at an angle of \SI{0}{\degree} relative to the sensor cable on the same plane.
Regarding the system, we varied two key factors. 
For signal coupling, we tested cable lengths beyond the \SI{16}{\centi\meter} default. 
For temperature effects on optically black pixels, we immersed the sensor in a temperature-controlled water bath, in which \SI{30}{\celsius} is the default. 
We also evaluated three additional sensors alongside the OV5647. 
See Table~\ref{tab:experiment-params} for details.

In most commercial image sensors, access to optically black pixels is not provided to end users. 
To address this limitation and enable empirical validation of our method, we construct optically black pixels by covering the active region of the sensor with opaque material, and using the \emph{raw, unprocessed} pixel values for analysis. 
As discussed earlier, the only structural distinction between optically black pixels and active pixels is the presence of a light-blocking shield; both are otherwise identical. 
Therefore, masking active pixels and using their raw values provides a practical approximation of genuine optically black pixels.
More discussion on the deployment feasibility of our proposed detection approach can be found in Section~\ref{sec:deployment_feasibility}.

\subsection{Visualization and Statistics}

\begin{figure}[t]
    \centering

    \begin{subfigure}{0.25\textwidth}
        \centering
        \includegraphics[width=\textwidth]{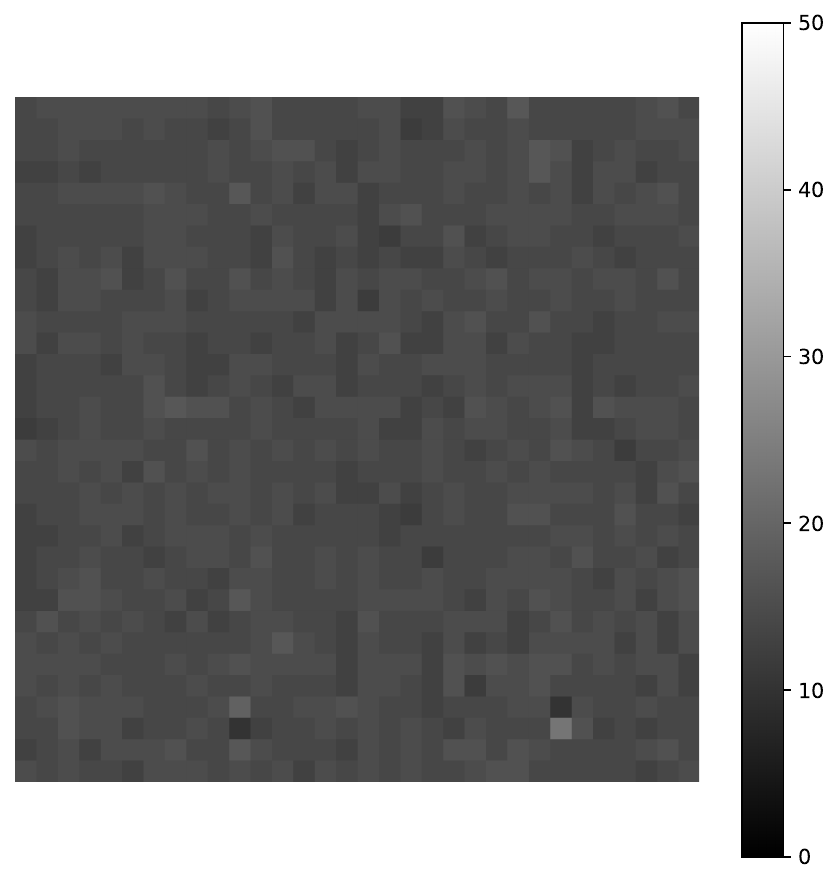}
        \caption{Without Attack.}
        \label{fig:visualization_wo_atk}
    \end{subfigure}
    \hfill
    \begin{subfigure}{0.25\textwidth}
        \centering
        \includegraphics[width=\textwidth]{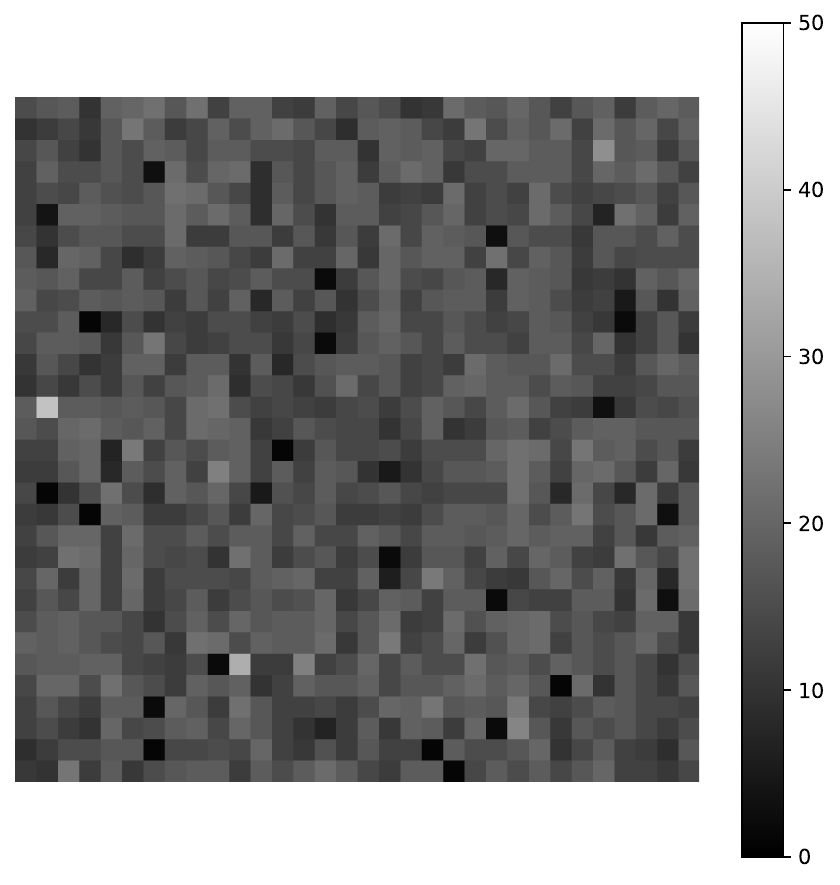}
        \caption{With Attack.}
        \label{fig:visualization_w_atk}
    \end{subfigure}
    \hfill
    \begin{subfigure}{0.4\textwidth}
        \centering
        \includegraphics[width=\textwidth]{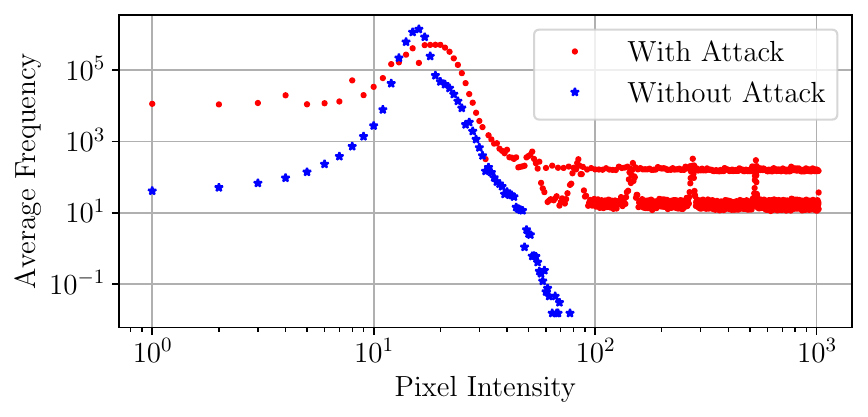}
        \caption{Pixel Distribution.}
        \label{fig:visualization_and_hist}
    \end{subfigure}
    \caption{A 16\(\times\)16 crop of optically black pixels and the averaged distributions of pixel intensities for the cases with and without attack.}
\vspace{-1em}
\end{figure}

We provide a visualized comparison of the pixel values in the optically black pixel region under two conditions: Figure~\ref{fig:visualization_wo_atk} without an attack, and Figure~\ref{fig:visualization_w_atk} with an attack.
As shown in the figures, in the absence of an attack, the optically black pixels remain relatively stable, exhibiting low variability and consistent intensity values. 
In contrast, when an ESIA is present, the optically black pixel values become significantly noisier, showing greater variation, with noticeable bright spots and distortions.

For each image, pixel intensities are extracted, and a histogram is computed. 
To obtain a representative distribution, the histograms from all images (under different attack conditions and system conditions as detailed above) are averaged, as shown in Figure~\ref{fig:visualization_and_hist}.
When there is no attack, no pixel intensity exceeds 100, and the peak appears around 16. However, when an attack occurs, the frequency of both low-intensity and high-intensity pixels increases significantly.

\begin{table}[t]
    \centering
    \caption{Comparison of statistical features between cases with and without attack (sorted by p-value).}
    \label{tab:statistical_features}
    \renewcommand{\arraystretch}{0.8}  
    \setlength{\tabcolsep}{4pt}  
    \adjustbox{max width=\textwidth}{ 
    \resizebox{0.6\columnwidth}{!}{%
    \begin{tabular}{lccccc}
        \toprule
        \textbf{Statistic} & \multicolumn{2}{c}{\textbf{Without Attack}} & \multicolumn{2}{c}{\textbf{With Attack}} & \textbf{p-value} \\
        \cmidrule(lr){2-3} \cmidrule(lr){4-5}
        & Mean & Std Dev & Mean & Std Dev & \\ 
        \midrule
        Std Dev   & 1.5   & 0.7   & 40.4  & 45.5  & 5e-22  \\
        Entropy   & 2.5   & 0.4   & 4.1   & 0.4   & 1e-21  \\
        Max       & 50.5  & 5.9   & 679.8 & 461.4 & 1e-21  \\
        Median    & 16.1  & 1.0   & 18.3  & 1.0   & 2e-20  \\
        Mean      & 15.9  & 0.9   & 23.8  & 11.1  & 3e-20  \\
        Kurtosis  & 1.8   & 0.4   & 688.3 & 1904.6 & 2e-19  \\
        Mode      & 16.3  & 0.9   & 18.3  & 1.9   & 1e-13  \\
        Skewness  & -0.1  & 0.1   & 12.8  & 17.5  & 1e-08  \\
        Min       & 0     & 0     & 0     & 0     & 1e0 \\
        \bottomrule
    \end{tabular}
    }
    }
\vspace{-2em}
\end{table}

Table~\ref{tab:statistical_features} presents a quantitative comparison between the ``With Attack'' and ``Without Attack'' scenarios across nine commonly used statistical features: mean, median, minimum, maximum, standard deviation, mode, entropy, skewness, and kurtosis. These features are computed from the pixel values of images containing only optically black pixels. 
For each case, we report the mean and standard deviation of each feature across all collected images.

To formally assess whether the distributions of optically black pixel values in the two cases are statistically different, we conduct the Mann–Whitney U test~\cite{macfarland2016mann} for each feature, under the null hypothesis ($H_0$) that the black pixels in both cases originate from the same distribution, and the alternative hypothesis ($H_1$) that they do not. 
We opt for the Mann–Whitney U test instead of a t-test, because the latter has a normality assumption that does not hold in this case.
We use a common confidence level of 95\%, corresponding to a significance level of 0.05.
As shown in the Table~\ref{tab:statistical_features}, the p-values for all features (except min) are extremely small, far below 0.05, leading us to reject $H_0$.
This shows that the optically black pixel distributions in the ``With Attack'' and ``Without Attack'' cases are significantly statistically different.

\subsection{Impacts of Different Parameters}

In this section, we further investigate how various parameters that are summarized in Table~\ref{tab:experiment-params} influence the impact of ESIA on optically black pixels.

\begin{figure*}[t]
    \centering
    \begin{subfigure}[t]{0.24\textwidth}
        \centering
        \includegraphics[width=\linewidth]{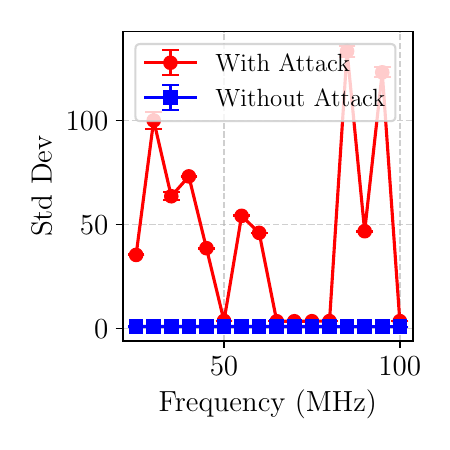}
        \caption{Frequency}
        \label{fig:stddev_vs_frequency}
    \end{subfigure}
    \hfill
    \begin{subfigure}[t]{0.24\textwidth}
        \centering
        \includegraphics[width=\linewidth]{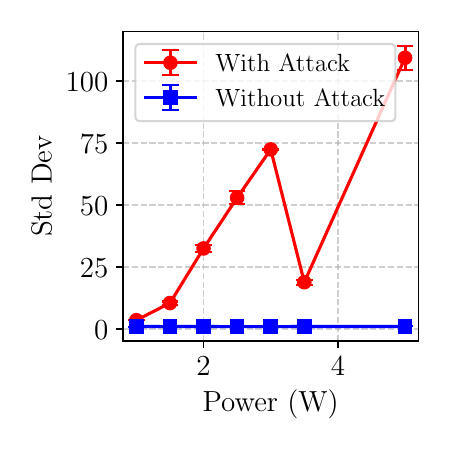}
        \caption{Power}
        \label{fig:stddev_vs_power}
    \end{subfigure}
    \hfill
    \begin{subfigure}[t]{0.24\textwidth}
        \centering
        \includegraphics[width=\linewidth]{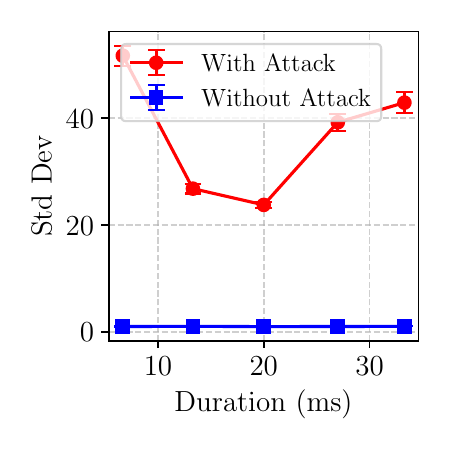}
        \caption{Duration}
        \label{fig:stddev_vs_duration}
    \end{subfigure}
    \hfill
    \begin{subfigure}[t]{0.24\textwidth}
        \centering
        \includegraphics[width=\linewidth]{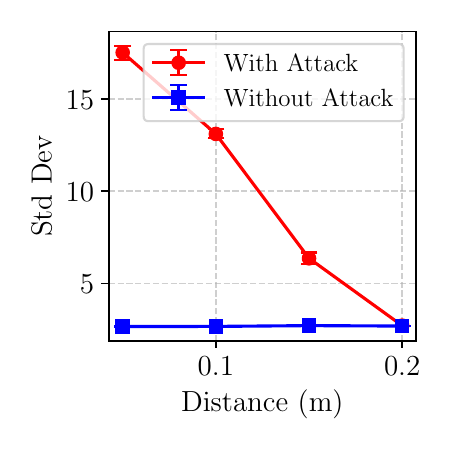}
        \caption{Distance}
        \label{fig:stddev_vs_distance}
    \end{subfigure}
    \caption{Standard deviation under different experimental conditions, including frequency, power, duration, and distance.}
    \vspace{-1.8em}
\end{figure*}

\textbf{Attack Frequency.}
Across all tested frequencies, we observed the presence of color strips in the active pixel regions, which indicates that the attack signal successfully couples into the imaging system over the entire frequency range.
To quantitatively assess the impact of the attacks on optically black pixels, we select the standard deviation (Std Dev) as an example hereafter regarding its lowest p-value. 
More visualization of other features, such as entropy and max, can be found in Appendix~\ref{apx:statistical_features_in_different_attack_conditions}. 

Figure~\ref{fig:stddev_vs_frequency} illustrates the behavior of standard deviation across the tested frequencies.
High separability is observed between the ``With Attack'' and ``Without Attack'' conditions throughout the frequency range. 
This suggests that optically black pixels are sensitive to injected signals across a wide frequency spectrum.
Notably, at specific frequencies, namely \SI{50}{\mega\hertz}, \SI{65}{\mega\hertz}, \SI{70}{\mega\hertz}, \SI{75}{\mega\hertz}, and \SI{80}{\mega\hertz}, the magnitudes of the feature values under attack are relatively lower compared to other attack frequencies, but, they are still statistically distinguishable from the non-attack baseline.
This phenomenon may be attributed to the efficiency of electromagnetic coupling between the injected signal and the circuitry of the image sensor. 
Given a fixed transmission power, the coupling effectiveness can vary with frequency due to factors such as circuit impedance and resonance characteristics. 
At the above-mentioned frequencies, it is likely that the injected signal couples less efficiently into the system (note that the attacks are successful), resulting in weaker disturbances in the optically black pixel regions.
In contrast, at other frequencies, the statistical features are significantly elevated under attack, indicating stronger coupling and more significant pixel disruptions.

\textbf{Attack Power.}
As illustrated in Figure~\ref{fig:stddev_vs_power}, the statistical feature demonstrates a general upward trend as the attack power increases. 
The increase in feature magnitudes can be attributed to the fact that higher signal power leads to more energy coupling into the internal circuitry, which in turn causes more measurable disturbances in the black pixels.

It is essential to point out that when the attack power is below \SI{2}{\watt}, the color strips in the active pixel region are barely noticeable. 
However, statistical features from the black pixels still reflect the presence of the attack, demonstrating their high sensitivity to weak attacks.
Interestingly, at \SI{3.5}{\watt}, there is a sudden drop in the feature value (i.e., Std Dev). 
This anomaly is likely due to instability in the RF amplification, which may have resulted in inconsistent output power or signal distortion at that specific setting.
Nevertheless, the \SI{3.5}{\watt} attack is still causing noticeable disturbance that is separable from the benign baseline values.
Beyond this dip, the features rise again at \SI{5}{\watt}, highlighting the overall trend that higher attack power leads to more severe impact on the optically black pixel statistics.

\textbf{Duration.}
As shown in Figure~\ref{fig:stddev_vs_duration}, the feature value decreases as duration is reduced from \SI{33.3}{\milli\second} to \SI{20}{\milli\second}, reaching a minimum at \SI{20}{\milli\second}, and then increases again as the duration becomes even shorter. 
This U-shaped trend can be explained by the interaction between the attack duration and the image sensor’s rolling shutter scan pattern.
At \SI{33.3}{\milli\second}, the signal is present throughout the entire frame capture, and thus most or all optically black pixel rows are uniformly affected, leading to higher feature values due to consistent but widespread disruption.
As the duration shortens, fewer rows are impacted, resulting in less overall variance.
However, when the duration becomes very short (e.g., \SI{6.7}{\milli\second}), the attack behaves like a brief pulse, creating high-intensity localized disturbances in a small subset of pixels, further leading to a bigger standard deviation of optical black pixel values.

\textbf{Attack Distance.}
As shown in Figure~\ref{fig:stddev_vs_distance}, the statistical feature exhibits a decreasing trend as the attack distance increases.
At shorter distances (e.g., \SI{0.05}{\meter}), the injected signal is stronger due to reduced path loss, leading to more energy being coupled into the internal circuitry. 
Consequently, the optically black pixel features show significantly higher values, indicating stronger disruption.
As the distance increases, the electromagnetic signal attenuates, resulting in weaker coupling into the system. This leads to a gradual reduction in the disturbance observed in the optically black pixels.
This behavior aligns with the inverse-square law of electromagnetic wave propagation, where signal strength decreases with the square of the distance from the source~\cite{friis1946note}.

When the distance between the attack antenna and the image sensor reaches \SI{0.2}{\meter}, no visible color strip is observed in the captured images. Correspondingly, the statistical features extracted from the optically black pixels under attack conditions converge with those from the non-attack baseline, showing no statistically significant difference.
Because no visible distortion occurs and optically black pixel statistics remain indistinguishable from the baseline, it is not necessary for the optically black pixels to detect or flag the presence of such weak signals. In other words, the statistical features are appropriately silent when the attack has no practical impact, avoiding false alarms.

\textbf{Attack Angle.}
\begin{figure*}[t]
    \centering
    \begin{subfigure}[t]{0.24\textwidth}
        \centering
        \includegraphics[width=\linewidth]{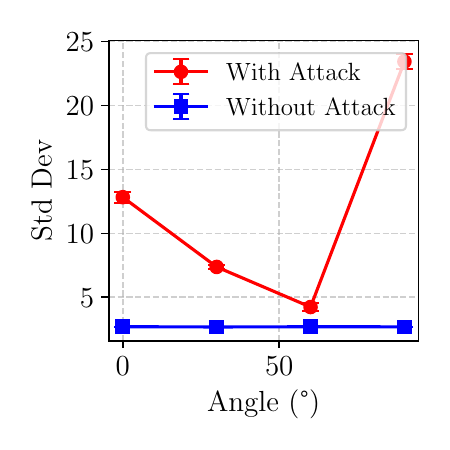}
        \caption{Attack angle}
        \label{fig:stddev_vs_angle}
    \end{subfigure}
    \hfill
    \begin{subfigure}[t]{0.24\textwidth}
        \centering
        \includegraphics[width=\linewidth]{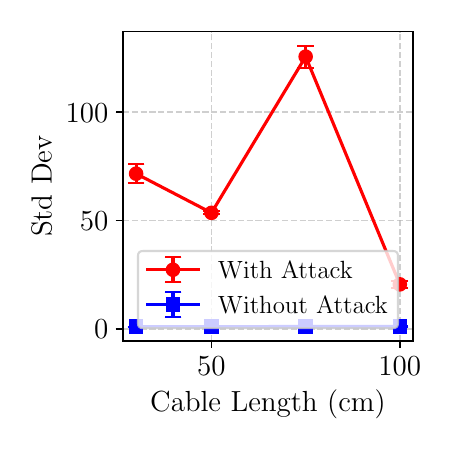}
        \caption{Cable length}
        \label{fig:stddev_vs_cable_length}
    \end{subfigure}
    \hfill
    \begin{subfigure}[t]{0.24\textwidth}
        \centering
        \includegraphics[width=\linewidth]{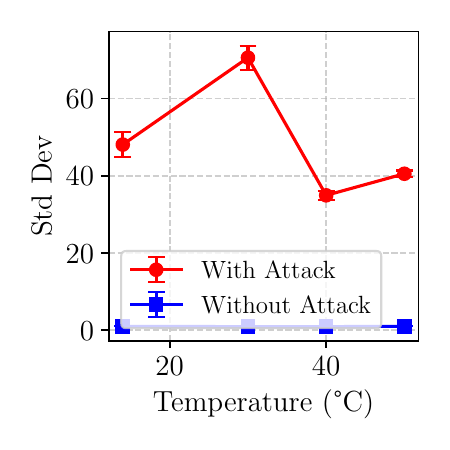}
        \caption{Temperature}
        \label{fig:stddev_vs_temperature}
    \end{subfigure}
    \hfill
    \begin{subfigure}[t]{0.24\textwidth}
        \centering
        \includegraphics[width=\linewidth]{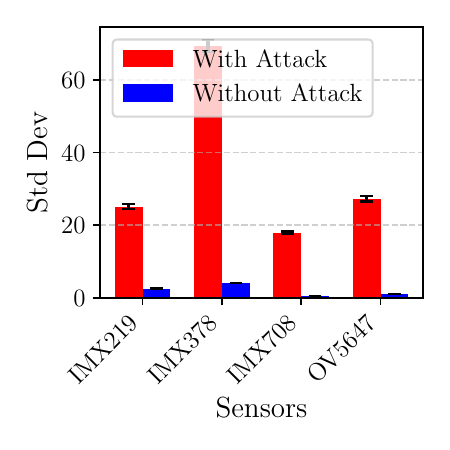}
        \caption{Sensor}
        \label{fig:stddev_vs_sensor}
    \end{subfigure}
    \caption{Standard deviation under different experimental conditions, including attack angle, cable length, temperature, and sensor model.}
    \vspace{-1.8em}
\end{figure*}
As shown in Figure~\ref{fig:stddev_vs_angle}, the standard deviation of the optically black pixels clearly distinguishes between the ``With Attack'' and ``Without Attack'' conditions across all angular configurations. 
The magnitude of the feature varies with angle:
at \SI{0}{\degree} and \SI{90}{\degree}, the feature exhibits relatively high values, suggesting strong coupling and noticeable impact on the optically black pixels;
at \SI{60}{\degree}, the feature reaches a local minimum, indicating a weaker disruption compared to other angles. 
This suggests that the coupling efficiency is reduced at this specific orientation.
However, even with changes in antenna orientation, the optically black pixels remain effective in capturing the presence of injected signals.

\textbf{Cable Length.}
As illustrated in Figure~\ref{fig:stddev_vs_cable_length}, at \SI{100}{\centi\meter}, the feature reaches its lowest values under attack conditions. 
This suggests that, at this length, the efficiency of signal coupling into the image sensor system is reduced compared to shorter cables.
In contrast, \SI{30}{\centi\meter} and \SI{75}{\centi\meter} cables exhibit higher feature values, indicating better coupling efficiency and, consequently, stronger disturbances in the optically black pixel region.
Despite this drop at \SI{100}{\centi\meter}, the feature under attack remains consistently higher than the non-attack baseline across all cable lengths. 
This indicates that the optically black pixel statistics still maintain sufficient separability to detect the presence of an attack.

\textbf{Temperature.}
As shown in Figure~\ref{fig:stddev_vs_temperature}, when no attack signal is present, the statistical features remain stable across all temperatures, showing only minimal fluctuations. 
This indicates that the optically black pixel region is not significantly affected by temperature alone, and that the baseline measurements are robust to thermal variation.
In contrast, when an attack is present, the statistical feature values remain consistently elevated compared to the non-attack baseline under all tested temperatures. 
This demonstrates that the injected signal continues to induce distinguishable disturbances in the optically black pixel region, regardless of thermal conditions.

\textbf{Sensor Model.}
We extended our experiments to include three additional commercially available image sensors. 
As shown in Figure~\ref{fig:stddev_vs_sensor}, for each sensor, feature values are significantly elevated under attack, indicating that optically black pixels are consistently affected by the injected signal.
Despite potential differences in internal architecture, layout, or analog front-end design, all tested sensors exhibit detectable attack-induced disruptions in their optically black pixel statistics.

Our findings confirm that optically black pixels potentially provide a robust and generalizable sensing mechanism for detecting ESIA.
The analysis of the detection performance is further presented and discussed in the following section.

\section{Analysis of Detection Performance}
\label{sec:analysis_of_detection_performance}

We now evaluate the performance of our proposed method quantitatively. 
This section presents an analysis using both raw pixel values and their statistical features.

\subsection{Metrics}
To quantify the effectiveness of this detection approach, we evaluate a threshold-based classification method using two commonly used metrics: Receiver Operating Characteristic – Area Under Curve (ROC-AUC, or simply denoted as AUC, hereafter)~\cite{huang2005using,sammut2011encyclopedia,schuckers2010receiver}, and Equal Error Rate (EER)~\cite{schuckers2010receiver,teh2016survey,conrad2016domain}.

AUC measures the trade-off between true positive rate (TPR) and false positive rate (FPR) across all possible threshold settings.
A higher AUC value (closer to 1) indicates better classification performance, with 1 representing perfect separation between the two classes.
An AUC of 0.5 indicates a performance no better than random guessing.

EER is the point at which the FPR equals the false negative rate (FNR, or 1-TPR). It reflects the threshold at which both types of classification errors are balanced.
A lower EER indicates better performance, as it represents fewer misclassifications overall.
An ideal classifier would achieve an EER of 0\%.

\subsection{Detection Using Raw Pixel Values}

Recalling the sensor readout architecture, image pixels are transmitted row by row, and the optically black pixels occupy a fixed set of columns across all rows. 
In our detection approach, we monitor the pixel values of these black columns for each row. 
If any row contains optically black pixel values that exceed a predefined threshold, we flag that frame as being under attack.

Based on the image data collected from the previous section, we first use 16 optically black pixel columns across all rows (in OV5647 image sensor), and compare the optically black pixel values per row to a threshold. 
If any value exceeds the threshold, the frame is classified as an attack. 
As a result, we achieve an AUC of 0.996 and an EER of 0.025, with the threshold found to be 21.69. 
This shows that even a simple rule based on raw pixel values can accurately detect injected signals.

\subsubsection{Impact of Number of Columns}

\begin{figure}[t]
    \centering
    \begin{subfigure}{0.43\textwidth}
        \centering
        \includegraphics[width=\textwidth]{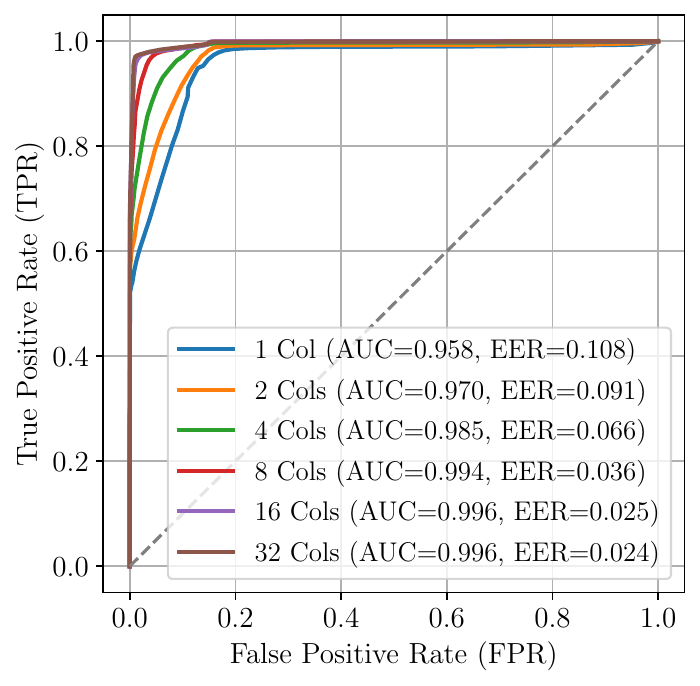}
        \caption{}
        \label{fig:roc_auc_diff_cols}
    \end{subfigure}
    \begin{subfigure}{0.43\textwidth}
        \centering
        \includegraphics[width=\textwidth]{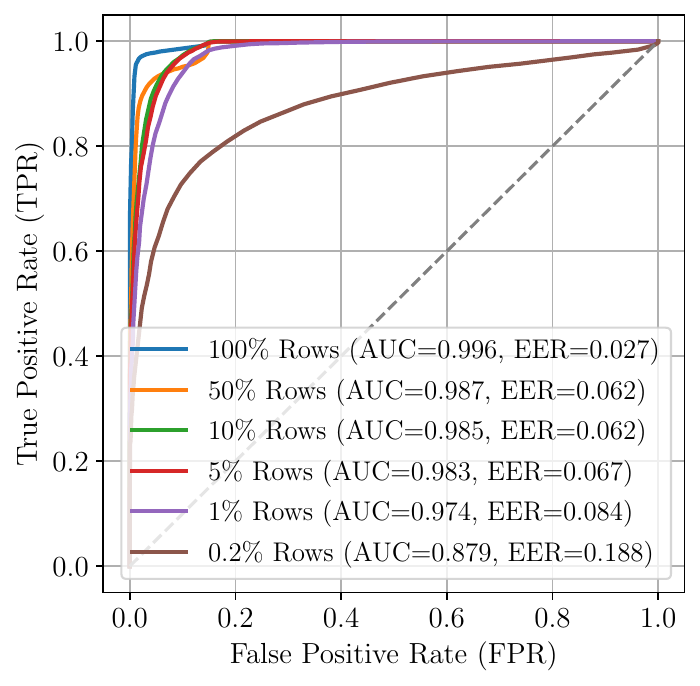}
        \caption{}
        \label{fig:roc_auc_diff_rows}
    \end{subfigure}
    \caption{ROC while using different numbers of (a) columns and (b) rows of optically black pixels.}
    \label{fig:roc_auc_diff_cols_and_rows}
\vspace{-2em}
\end{figure}

We investigate how the number of optically black pixel columns affects detection performance. Specifically, we vary the number of columns from 1, 2, 4, 8, 16, to 32, and apply the same threshold-based method. Results are shown in Figure~\ref{fig:roc_auc_diff_cols}.

We observe that as the number of columns increases, performance improves.
Even with just 1 column, the AUC reaches 0.958 and EER is 0.108, suggesting that a small number of optically black pixels already carry useful information about the presence of an attack.
With 4 or 8 columns, performance improves significantly ($\text{AUC} > 0.98$, $\text{EER} < 0.07$).
At 16 or 32 columns, performance saturates ($\text{AUC} \approx 0.996$, $\text{EER} \approx 0.025$), indicating that wider spatial coverage enhances stability and robustness.

\subsubsection{Impact of Number of Rows}

We further explore how the number of rows used per frame affects detection accuracy (keeping the number of columns to 16). 
Rather than utilizing all available rows, we randomly sample a subset of rows at various percentages: 50\%, 10\%, 5\%, 1\%, and 0.2\%. 
For each configuration, we apply the same threshold-based detection rule, comparing raw optically black pixel values to a predefined threshold, and evaluate the resulting performance. 
The results are presented in Figure~\ref{fig:roc_auc_diff_rows}, which shows the ROC curves for each row sampling rate.

Using 50\% or even 10\% of the rows maintains strong detection performance, with AUC above 0.985 and EER values below 0.062. 
This may suggest that analyzing the entire frame is not necessary for reliable detection, and row reduction is possible without sacrificing accuracy too much.
Remarkably, even when only 1\% of the rows are sampled, the system still achieves good performance ($\text{AUC} = 0.974$, $\text{EER} = 0.084$).
However, when the number of sampled rows is reduced to just 0.2\% (i.e., 3 or 4 rows of pixels per frame), performance degrades noticeably ($\text{AUC} = 0.879$, $\text{EER} = 0.188$). 
This may indicate that overly sparse sampling may fail to capture the temporal or spatial effects of the injected signal, leading to missed detections.

\subsection{Detection Using Statistical Features}

Beyond directly comparing individual optically black pixel values, we further examine the use of statistical features computed over optically black pixels for threshold-based detection. By aggregating values across rows and columns, these features can enhance robustness to noise while preserving computational simplicity. In this study, we consider several common statistical measures: mean, median, maximum, standard deviation, skewness, kurtosis, entropy, and mode.

\begin{figure}[t]
    \centering
    \begin{subfigure}{0.43\textwidth}
        \centering
        \includegraphics[width=\textwidth]{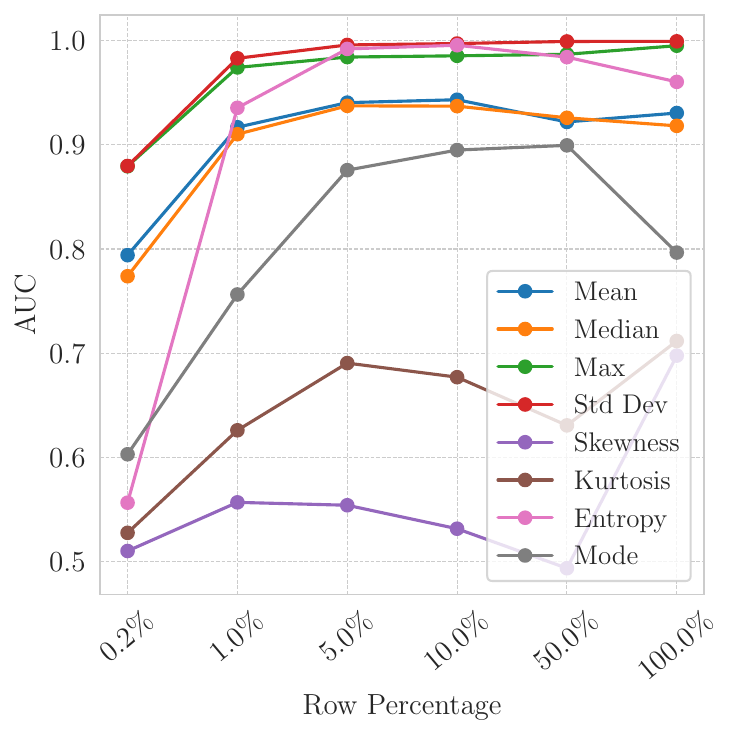}
        \caption{}
    \end{subfigure}
    \begin{subfigure}{0.43\textwidth}
        \centering
        \includegraphics[width=\textwidth]{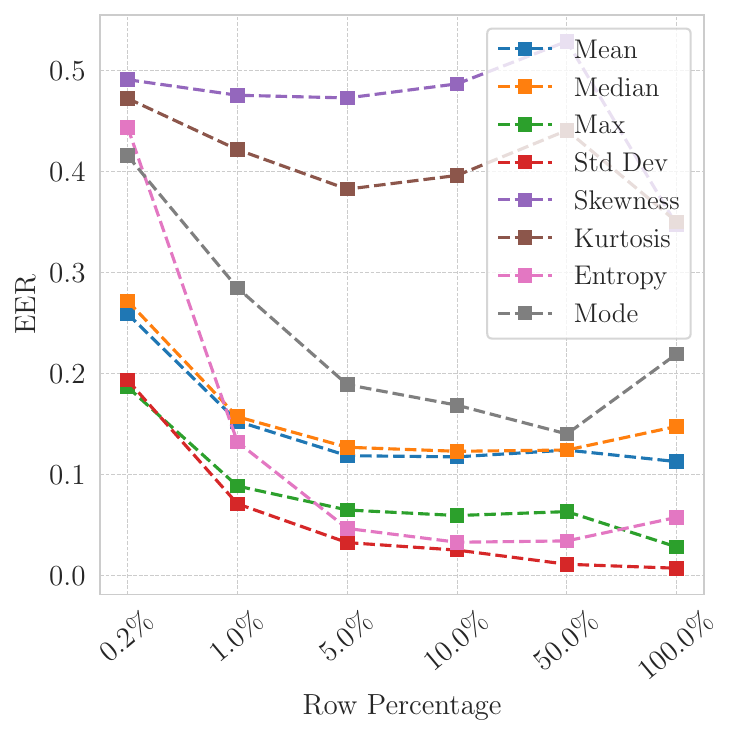}
        \caption{}    
    \end{subfigure}
    \caption{(a) AUC and (b) EER while using different number of rows of optically black pixels.}
    \label{fig:diff_features_auc_eer_vs_row}
\vspace{-2em}
\end{figure}

We fix the number of optically black pixel columns to 16 and vary the number of rows used by randomly sampling different percentages of the frame: 0.2\%, 1\%, 5\%, 10\%, 50\%, and 100\%. For each feature, we apply a simple threshold rule and evaluate detection performance using AUC and EER. 
The results are shown in Figure~\ref{fig:diff_features_auc_eer_vs_row}, where subfigure (a) reports AUC values and subfigure (b) shows the corresponding EER values.

Among all features evaluated, standard deviation, maximum, and entropy consistently demonstrate the best detection performance. 
Even at 1\% of rows, these three features achieve high AUC values (above 0.98) and low EERs (below 0.1). 
As the number of rows increases, their performance further improves, with AUC approaching 1.0 and EER dropping below 0.03. 
This indicates that these features are highly sensitive to the signal perturbations introduced by electromagnetic injection, and remain robust even under limited data conditions.

In contrast, other features such as mean, median, skewness, kurtosis, and mode exhibit weaker performance.
Mean and median offer moderate detection accuracy when a sufficient number of rows are available (e.g., above 10\%), but are less reliable under other conditions. 
Skewness and kurtosis, which are higher-order moments of the pixel distribution, are especially sensitive to sample size. 
Their performance improves noticeably with more data, but remains worse than other features.


\subsection{Performance Across Different Image Sensors}

From the above analysis, we conclude that using standard deviation as the classification feature yields the most effective results among all statistical measures. 
It demonstrates both high sensitivity to injected signals and strong robustness under varying sampling conditions.
Building on this observation, we further investigate how this rule-based detection method performs across different image sensors. 
Doing so aims to assess the generalization and robustness of the approach when applied to data captured by sensors with varying characteristics, such as resolution, noise profile, and manufacturing differences.

We test its performance across four different image sensors: OV5647, IMX219, IMX708, and IMX378. 
The results are summarized in Table~\ref{tab:sensor_comparison}, which reports the number of active pixels and optically black pixels, the detection threshold for each sensor, and the corresponding AUC and EER values.
For each sensor, the detection threshold is selected such that the resulting EER is achieved.

As shown in the table, the method achieves consistently high performance across all sensor types. The AUC values range from 0.980 to 0.996, while the EER values remain below 0.05 in all cases.
Specifically, the OV5647 sensor achieves the highest AUC of 0.996 with an EER of 0.027. The IMX708 and IMX378, despite using a smaller optically black pixel region (only 4 columns), also perform well, achieving AUCs above 0.99 and EERs below 0.012. 
This suggests that the method remains effective even when fewer optically black pixels are available, provided that the standard deviation signal is strong enough.

The variation in the threshold values (ranging from 3.026 to 5.829) reflects differences in sensor-specific noise floors and analog signal behavior. 
For example, the IMX708 requires a higher threshold of 5.829, likely due to a higher baseline fluctuation in its optically black pixel readings, possibly caused by internal gain or readout characteristics. 
Nevertheless, once properly calibrated, the threshold-based method maintains strong detection capability across different image sensors.
The detection threshold used in the proposed detection approach is software-defined and can be adjusted dynamically. 
This flexibility enables the system to adapt to varying electromagnetic environments or sensor-specific characteristics that may drift over time due to factors such as aging. 
For example, in more complex or noisy environments, the threshold can be adjusted to maintain detection performance while minimizing the EER. 

\begin{table}[t]
\centering
\caption{Performance Comparison between Sensors}
\label{tab:sensor_comparison}
\resizebox{0.6\columnwidth}{!}{%
\begin{tabular}{lllll}
\hline
Sensors         & OV5647           & IMX219           & IMX708           & IMX378           \\ \hline
Active Pixels   & 2592$\times$1944 & 3280$\times$2464 & 4608$\times$2592 & 4056$\times$3040 \\
Black Pixels    & 16$\times$1944   & 16$\times$2464   & 4$\times$2592    & 4$\times$3040    \\
Threshold       & 3.026            & 4.596            & 5.829            & 4.729            \\
AUC$\uparrow$   & 0.996            & 0.980            & 0.992            & 0.991            \\
EER$\downarrow$ & 0.027            & 0.042            & 0.009            & 0.012            \\ \hline
\end{tabular}%
}
\end{table}
\section{Discussion}
\label{sec:discussion}

In this section, we reflect on the broader implications of our proposed detection method, compare it with alternative approaches, and discuss its deployment feasibility. 

\subsection{Temporal Analysis of Detection Failures}

There is an inherent attacker trade-off: weaker attacks are stealthier but less effective, while stronger attacks are more disruptive yet easier to detect via optically black pixel statistics. This makes ``stealthy yet effective'' attacks difficult in practice.
With an EER of 0.027 on OV5647, both the FNR and FPR are 2.7\%. 
Imagining in a common 30 frames per second setting, sustaining manipulation requires fooling every frame; the probability of bypassing all 30 frames in one second is \(
P_{\text{bypass}} = (0.027)^{30} \approx 8.72 \times 10^{-48}
\), which is negligible.

The chance of 30 consecutive false positives is equally small, \(
P_{\text{false\_alert}} = (0.027)^{30} \approx 8.72 \times 10^{-48}
\), so prolonged false alerts are implausible. 
Single-frame false positives can be suppressed with simple temporal filters (e.g., majority voting). 
Overall, despite a nonzero per-frame error, sustained undetected attacks and persistent false alarms are vanishingly unlikely.

\subsection{Comparison with Machine Learning (ML) Classifiers}
\label{sec:comparison_with_ml_and_dl_classifers}

Our method is simple and highly accurate. We also tested whether ML methods could improve performance using a Random Forest (RF) classifier on statistical features and a Convolutional Neural Network (CNN) on raw image slices. Details are provided in Appendix~\ref{apx:detection_using_ml_dl}.
Both RF and CNN achieve near-perfect results (Table~\ref{tab:dl_results} in Appendix~\ref{apx:detection_using_ml_dl}). 
However, these methods require much higher computational cost: particularly, CNN inference takes 0.1--0.2\,s per frame, while our method takes only 0.000135\,s per frame on a Raspberry Pi. 
Therefore, our method remains preferable for real-time, edge, and mobile deployment.

\subsection{Attack Distance}
Long-range attacks (e.g., \SI{10}{\meter} as shown in prior work~\cite{brokenwire}) require higher transmit power; under free-space assumptions, doubling distance typically demands roughly fourfold power to maintain the same coupling at the target~\cite{friis1946note}. 
Our experiments were constrained by local emission regulations and the amplifier’s power limit. Crucially, however, our detection method is agnostic to attacker distance: if the adversary induces any disturbance measurable by the image sensor, our method will detect it, regardless of distance.

\subsection{Deployment Feasibility}
\label{sec:deployment_feasibility}

There are two viable deployment paths:

\textbf{On-ISP Integration:} 
Image sensor manufacturers can natively implement the proposed monitoring logic within the Image Signal Processing (ISP) firmware, using the already available optically black pixel readings and reporting detection results or alerts directly to the system.

\textbf{Vendor-Enabled Access:} Alternatively, image sensor manufacturers can provide an API or firmware update that exposes optically black pixel values to system integrators or end users, enabling them to implement our lightweight detection algorithm at the application or middleware layer.

Both options are technically feasible and require minimal hardware changes. 
It is essential to highlight that the deployment feasibility of our detection method has been confirmed by a world-leading image sensor manufacturer and a leading vendor in Asia.
The low implementation overhead and significant security benefits provide strong incentives for vendors to adopt such features in future products, especially for mission-critical application scenarios.

\subsection{Future Work}

Future work could explore how CMOS sensor aging affects detection performance by periodically evaluating the sensor (for example, once a month) while continuously using it for recording between evaluations. 
Another direction would be to investigate the applicability of optically-black-pixel-based detection, which is also briefly mentioned by Ren et al.~\cite{ren2025ghostshot}, to CCD sensors. It is essential to note that, on CMOS, ESIA attacks' impacts are row‑based, so the optically black pixels in each row can directly signal an attack; on CCD, however, ESIA targets individual pixels, meaning the row‑oriented black‑pixel method proposed in this work may not directly apply and requires dedicated validation.

\section{Related Work}
\label{sec:related_work}

Before malicious electromagnetic signals reach a victim, they must first disturb environmental electromagnetic levels. 
One approach of defense is thus to monitor the electromagnetic environment for abnormal activity. 
For example, Adami et al.~\cite{adami2011hpm,adami2014hpm} developed low-cost electromagnetic detectors capable of identifying the attacks based on frequency-domain signatures.
Kune et al.~\cite{kune2013ghost} proposed using a reference conductor embedded in the system to capture electromagnetic waveforms and estimate ambient interference. 
Xu et al.~\cite{xu2021inaudible} suggested deploying parallel circuits specifically designed to detect modulated RF signals. 

When signal injection succeeds in entering the target system, detection can occur at the signal or application level. 
Several works embed secret modulation patterns or watermarks into sensor signals, which are verified to detect tampering~\cite{zhang2020detection,kohler2022signal}.
Researchers have also explored the use of physical-layer fingerprints, which have been applied to protect voice interfaces~\cite{kasmi2015iemi,fokkens2021machine}, industrial IoT systems~\cite{fang2022detection,wang2022deep}.
Armengol et al.~\cite{armengol2023brain} showed that phase shift anomalies in brain-computer interfaces can serve as indicators of interference. 
Zhang and Rasmussen~\cite{zhang2022detection}, Tu et al.~\cite{tu2021transduction}, and Kasmi et al.~\cite{kasmi2014autonomous} employed redundant or auxiliary sensors (or channels) to identify inconsistencies in sensor readings. 
Forecasting-based approaches, such as those by Zhang et al.~\cite{zhang2020observer} and Muniraj et al.~\cite{muniraj2019detection}, use historical data to predict future sensor outputs and flag deviations. 
Other systems monitor application-specific behaviors, such as touchscreen interaction intervals~\cite{wang2022ghosttouch}, LiDAR point distributions~\cite{bhupathiraju2023emi}, sensor signal responses~\cite{jiang2022wight}, or unpredictable changes in the wireless channel~\cite{rezaee2025ripple} to detect attack-induced anomalies.

\textbf{Differences between Our Method and Others.}
Our method addresses a domain where prior defenses struggle. 
Earlier methods focus on simple analog sensors (e.g., load cells, thermistors, microphones) with low-dimensional outputs, often relying on matched dummy circuits or duplicated signal paths.
However, these approaches are costly and complex to implement with image sensors. 
One potential reason is that image sensors typically involve millions of pixels and channels, 
which makes it impractical to faithfully replicate the same behavior on two sensors~\cite{tu2021transduction}. Any minor impedance, layout, or filtering mismatches between the main and dummy can cause false positives or false negatives in attack detection. 
In contrast, our method exploits an internal, scene-independent reference for detection: optically black regions that should remain quiescent regardless of content or task, avoiding dependence on application-layer patterns or ML models that require training and tuning.

\section{Conclusion}
\label{sec:conclusion}

In this work, we present a lightweight and effective detection method for electromagnetic signal injection attacks (ESIA) targeting CMOS image sensors. 
Our approach leverages optically black pixels, which are non-illuminated sensor elements traditionally used for calibration, as a monitoring channel to detect anomalies caused by malicious signal injection.
Through theoretical modeling, statistical analysis, and extensive experimentation, we demonstrate that optically black pixel values exhibit significant and consistent deviations under attack across various sensor models and attack parameters. 
Our detection method achieves high accuracy with minimal false positives, and offers a low-cost, practical, and effective defense mechanism to detect ESIAs targeting image sensors.

\section*{Acknowledgement}
This work was in part supported by Hong Kong RGC
Project (PolyU15227825), GRF15210023, The Hong Kong Polytechnic University under
grants P0059492 and P0053067, as well as grants from the CUHK IE department (project code: GRF/23/SYC, and GRF/24/SYC).

\bibliographystyle{splncs04}
\bibliography{main}

\appendix
\section{Detection Using Machine/Deep Learning}
\label{apx:detection_using_ml_dl}

In this appendix, we provide detailed information about the configurations and implementation of the machine learning (ML) and deep learning (DL) methods that can be applied to detect electromagnetic signal injection attacks (ESIA) on image sensors.
We implemented two classifiers, i.e., Random Forest (RF) on statistical features and a Convolutional Neural Network (CNN) on raw image slices, running on an R9 9950X CPU and RTX 4090 GPU.

\subsection{Dataset}
Using the setup in Section~\ref{sec:experiements}, we collected a dataset consisting of 90 images of street scenes subjected to ESIA, with careful manual annotation to identify three distinct attack consequences~\cite{jiang23glitchhiker}:

\begin{itemize}  
    \item \textbf{Hiding Attacks}: Cases where labeled objects (persons or vehicles) were not detected by the object detection models
    \item \textbf{Creating Attacks}: Instances where objects were falsely detected in regions without valid labels
    \item \textbf{Altering Attacks}: Situations where correctly detected objects were misclassified (e.g., a person identified as a vehicle)
\end{itemize}

For each image, we confirmed that at least one of our three benchmark object detection models exhibited these attack consequences. The models were chosen to represent the major families of modern object detectors:
\begin{itemize}
    \item \textbf{Co-DETR}~\cite{zong2023detrs}: A state-of-the-art transformer-based detector representing the latest advances in attention-based architectures
    \item \textbf{Faster R-CNN}~\cite{ren2015faster}: The canonical two-stage detector that has been widely adopted in industrial applications
    \item \textbf{YOLOv12}~\cite{tian2025yolov12}: The newest iteration of the single-stage YOLO family, known for its exceptional speed-accuracy tradeoff
\end{itemize}

\subsection{Training Set Construction}
For each of the 90 images (each with dimensions 2592$\times$1944 pixels), we conducted the following steps:

\begin{itemize}
    \item We extracted image slices from two distinct regions: the active pixel region and the optically black pixel region (covered by opaque materials).
    \item For each region, we generated slices with heights of 100 pixels and varying widths (1, 4, 8, and 16 pixels).
    \item We created 100 unique slice positions per image per width configuration, resulting in 100$\times$90 = 9,000 slices total before cross-validation splitting.
\end{itemize}

Such a slicing strategy was designed to test our method's sensitivity to different amounts of contextual information while maintaining computational efficiency. 
Narrower slices (e.g., 1-4 pixels) represent highly constrained scenarios where minimal image data is available, while wider slices (8-16 pixels) provide more context but require greater processing resources.

\subsection{Feature Extraction Methods}
We implemented and compared two distinct feature extraction approaches to thoroughly evaluate our detection methodology:

\subsubsection{Statistical Feature Extraction}
Our first approach computed seven fundamental statistical measures from each image slice, including mean, standard deviation, median, minimum, maximum, skewness, and kurtosis.
These statistics form a 7-dimensional feature vector that captures basic characteristics of the pixel intensity distribution. 

\subsubsection{Deep Learning-based Feature Extraction}
For our second approach, we designed a custom convolutional neural network (CNN) architecture specifically for this detection task. The network comprises:

\begin{itemize}
    \item \textbf{Input Layer}: Accepts grayscale image slices of varying widths (1-16 pixels) with fixed 100-pixel height
    \item \textbf{Convolutional Blocks}:
    \begin{itemize}
        \item First block: 4 filters of 3×1 kernels with ReLU activation and same padding.
        \item Second block: 32 filters of 3×1 kernels with ReLU, followed by 2×1 max pooling when input width $\geq$ 2.
        \item Third block: 64 filters of 3×1 kernels with ReLU, with 2×1 pooling when width $\geq$ 4.
        \item Fourth block: 128 filters of 3×1 kernels with ReLU, followed by global average pooling.
    \end{itemize}
    \item \textbf{Feature Projection}: A dense layer with 64 units and ReLU activation reduces the pooled features to a compact representation.
    \item \textbf{Output}: A single sigmoid unit for binary classification (attack/no attack).
\end{itemize}

The network's progressive pooling architecture dynamically adapts to different input widths while maintaining consistent feature dimensionality. 

\subsection{Training and Evaluation Methodology}

The Random Forest classifier was chosen as the final predictor based on its demonstrated robustness to feature scale variations, inherent feature importance analysis capabilities, and computationally efficient training and inference suitable for production environments.
We implemented 10-fold cross-validation with 8,100 training slices (90\% of 9,000) and 900 test slices per fold, assessing performance through precision, recall, and F1-score metrics for comprehensive detection capability analysis.
The results are shown in Table~\ref{tab:dl_results}.
It can be observed that ML/DL can achieve almost perfect detection performance.

Note that our primary focus is on optically black pixels due to their content-invariant baseline, but our experiments indicate that ML/DL models can also detect attacks using active pixel regions, albeit with increased risk of scene-dependent false positives as discussed previously. 
In practice, a hybrid approach that monitors both black and active pixels could possibly further reduce detection errors, especially for sensors with limited black pixel regions or in scenes lacking sufficient dark areas. 
Designing adaptive, ensemble-based detection strategies is a promising avenue for future research.

\subsection{Results}

Both RF and CNN achieve near-perfect performance, as shown in Table~\ref{tab:dl_results}, ``Black'' labels.
The results present the Precision (P), Recall (R), F-scores (F1), and Accuracy (Acc).
The CNN reaches F-scores of 1 even with just one column of optically black pixels, indicating learned features reliably detect electromagnetic injection.
However, ML/DL incurs substantial cost, i.e., a powerful CPU/GPU. 
Particularly, CNN inference takes 0.1 - 0.2 s per frame due to convolution/pooling. 
Our rule-based method processes a frame in 0.000135 s on the Raspberry Pi, yielding orders-of-magnitude speedups and making it preferable for real-time, edge, and mobile deployments.

We also tested detection using active pixel regions. 
RF and CNN models trained on active pixels achieved F-scores = 1 in 5-fold cross-validation, 
showing feasibility and performance comparable to optically black pixels.
However, unlike optically black pixels with a stable, near-zero baseline, active pixels have no fixed ground truth and vary with scene content, lighting, textures, and motion. 
These scene-dependent confounders can mimic adversarial patterns and cause false positives~\cite{guesmi2023physical}. 
Consequently, ML/DL on active pixels needs substantially more diverse data to generalize across environments.

\begin{table}[h]
\vspace{-2em}
\centering
\caption{Comparison of detection performance between the active image region and the optically black pixel region while using ML classifiers.}
\label{tab:dl_results}
\resizebox{0.5\columnwidth}{!}{%
\begin{tabular}{lllllll}
\hline
\textbf{Model} &
\makecell[l]{\textbf{Active/Black} \\ \textbf{Pixels}} &
\makecell[l]{\textbf{Pixel} \\ \textbf{Width}} &
\textbf{P} &
\textbf{R} &
\textbf{F1} &
\textbf{Acc} \\ \hline

\multirow{8}{*}{CNN} &
  Active &
  1 &
  1.000 &
  1.000 &
  1.000 &
  1.000 \\
 & Active & 4  & 1.000 & 1.000 & 1.000 & 1.000 \\
 & Active & 8  & 1.000 & 1.000 & 1.000 & 1.000 \\
 & Active & 16 & 1.000 & 1.000 & 1.000 & 1.000 \\
 & Black  & 1  & 1.000 & 1.000 & 1.000 & 1.000 \\
 & Black  & 4  & 1.000 & 1.000 & 1.000 & 1.000 \\
 & Black  & 8  & 1.000 & 1.000 & 1.000 & 1.000 \\
 & Black  & 16 & 1.000 & 1.000 & 1.000 & 1.000 \\ \hline

\multirow{8}{*}{\makecell{Random\\Forest}} &
  Active &
  1 &
  0.999 &
  0.990 &
  0.995 &
  0.995 \\
 & Active & 4  & 1.000 & 1.000 & 1.000 & 1.000 \\
 & Active & 8  & 1.000 & 1.000 & 1.000 & 1.000 \\
 & Active & 16 & 1.000 & 1.000 & 1.000 & 1.000 \\
 & Black  & 1  & 1.000 & 1.000 & 1.000 & 1.000 \\
 & Black  & 4  & 1.000 & 1.000 & 1.000 & 1.000 \\
 & Black  & 8  & 1.000 & 1.000 & 1.000 & 1.000 \\
 & Black  & 16 & 1.000 & 1.000 & 1.000 & 1.000 \\ \hline
\end{tabular}%
}
\vspace{-1.8em}
\end{table}

\section{Statistical Features in Different Attack Conditions}
\label{apx:statistical_features_in_different_attack_conditions}

\begin{figure}[h]
    \centering
    \begin{subfigure}{0.24\textwidth}
        \centering
        \includegraphics[width=\textwidth]{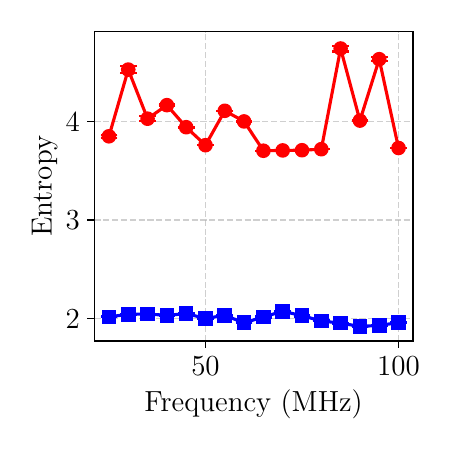}
        \caption{Entropy}
    \end{subfigure}\hfill
    \begin{subfigure}{0.24\textwidth}
        \centering
        \includegraphics[width=\textwidth]{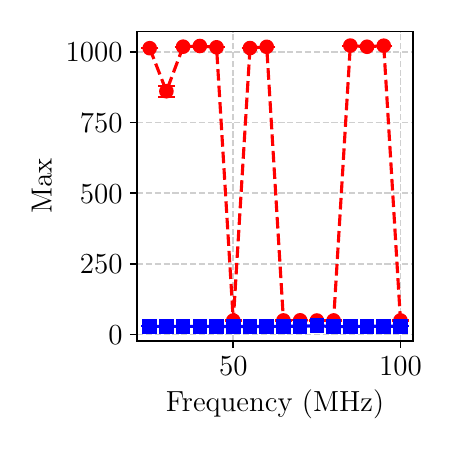}
        \caption{Max}
    \end{subfigure}\hfill
    \begin{subfigure}{0.24\textwidth}
        \centering
        \includegraphics[width=\textwidth]{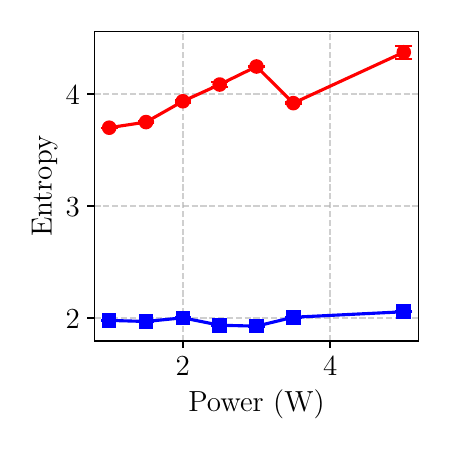}
        \caption{Entropy}
    \end{subfigure}\hfill
    \begin{subfigure}{0.24\textwidth}
        \centering
        \includegraphics[width=\textwidth]{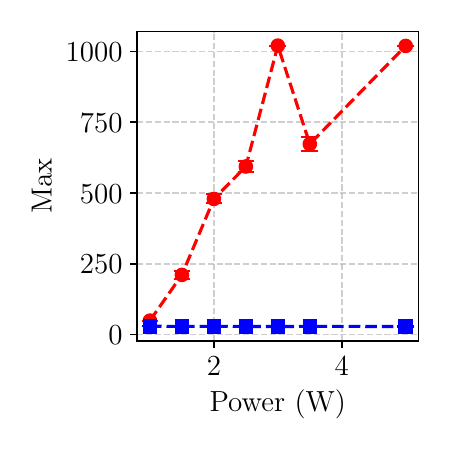}
        \caption{Max}
    \end{subfigure}
    \caption{ (a) and (b): Frequency; and (c) and (d): Power. The red dotted line represents that an attack is happening, while the blue squared line represents a benign (no attack) case.}
    \label{fig:entr-max_freq_power}
\end{figure}

\begin{figure}[h]
    \centering
    \begin{subfigure}{0.24\textwidth}
        \centering
        \includegraphics[width=\textwidth]{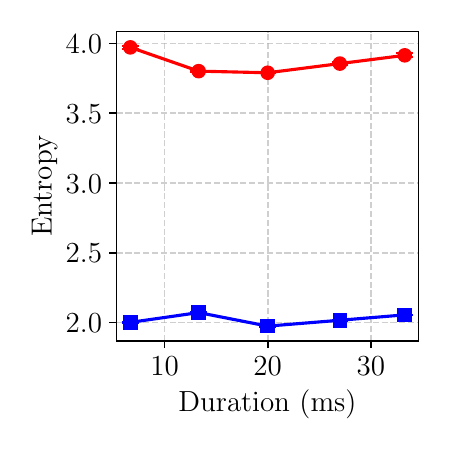}
        \caption{Entropy}
    \end{subfigure}\hfill
    \begin{subfigure}{0.24\textwidth}
        \centering
        \includegraphics[width=\textwidth]{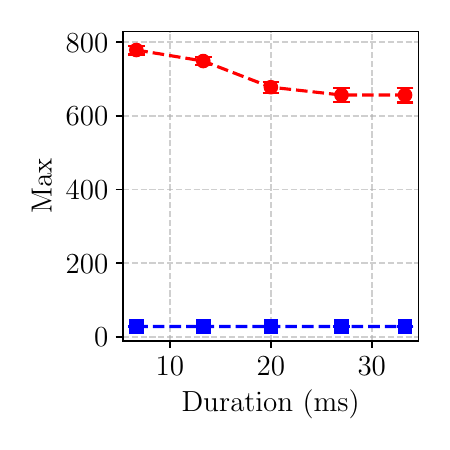}
        \caption{Max}
    \end{subfigure}\hfill
    \begin{subfigure}{0.24\textwidth}
        \centering
        \includegraphics[width=\textwidth]{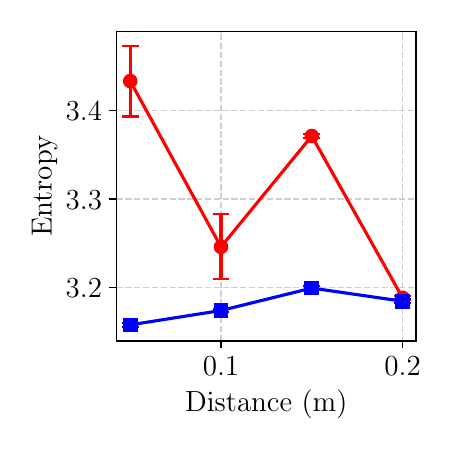}
        \caption{Entropy}
    \end{subfigure}\hfill
    \begin{subfigure}{0.24\textwidth}
        \centering
        \includegraphics[width=\textwidth]{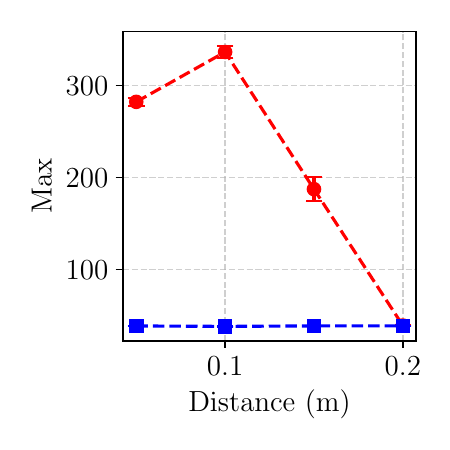}
        \caption{Max}
    \end{subfigure}
    \caption{(a) and (b): Duration; and (c) and (d): Distance.}
    \label{fig:entr-max_dur_dist}
\end{figure}

\begin{figure}[h]
    \centering
    \begin{subfigure}{0.24\textwidth}
        \centering
        \includegraphics[width=\textwidth]{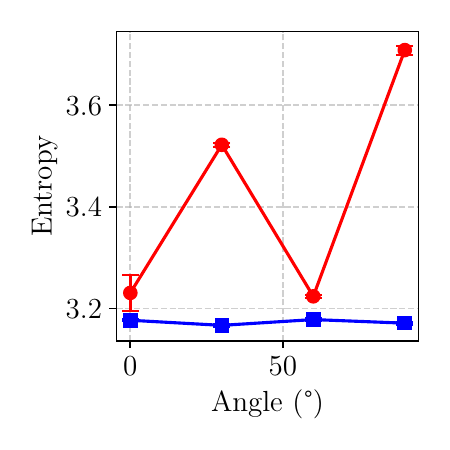}
        \caption{Entropy}
    \end{subfigure}\hfill
    \begin{subfigure}{0.24\textwidth}
        \centering
        \includegraphics[width=\textwidth]{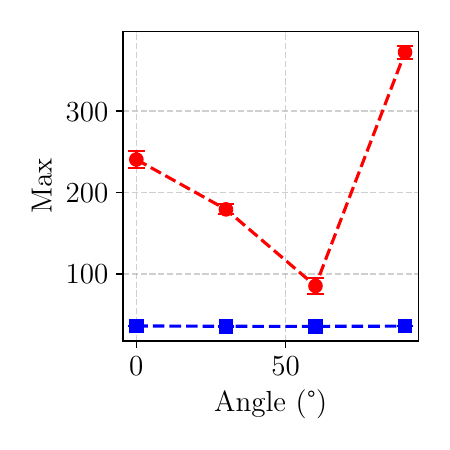}
        \caption{Max}
    \end{subfigure}\hfill
    \begin{subfigure}{0.24\textwidth}
        \centering
        \includegraphics[width=\textwidth]{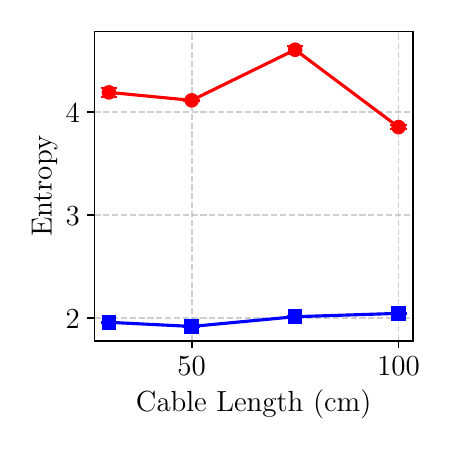}
        \caption{Entropy}
    \end{subfigure}\hfill
    \begin{subfigure}{0.24\textwidth}
        \centering
        \includegraphics[width=\textwidth]{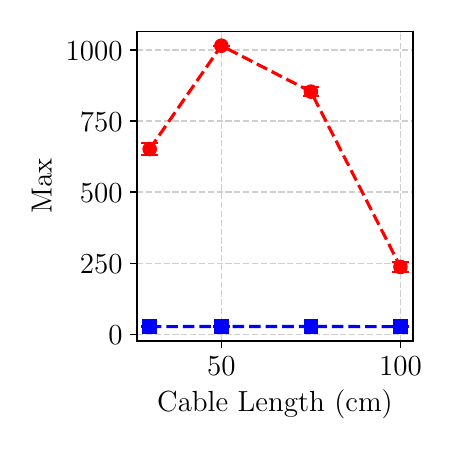}
        \caption{Max}
    \end{subfigure}
    \caption{(a) and (b): Attack Angle; and (c) and (d): Cable Length.}
    \label{fig:entr-max_angle_cable}
\end{figure}

\begin{figure}[h]
    \centering
    \begin{subfigure}{0.24\textwidth}
        \centering
        \includegraphics[width=\textwidth]{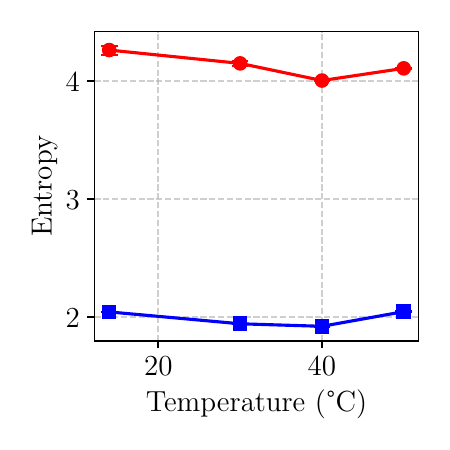}
        \caption{Entropy}
    \end{subfigure}\hfill
    \begin{subfigure}{0.24\textwidth}
        \centering
        \includegraphics[width=\textwidth]{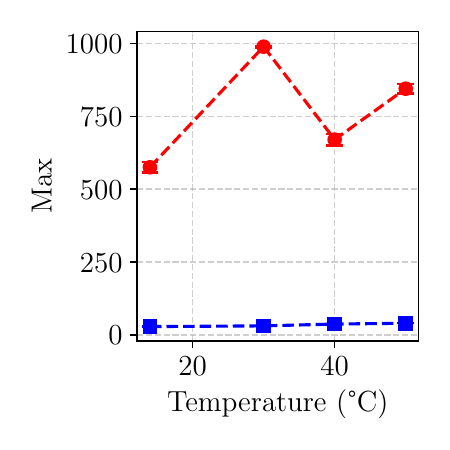}
        \caption{Max}
    \end{subfigure}\hfill
    \begin{subfigure}{0.24\textwidth}
        \centering
        \includegraphics[width=\textwidth]{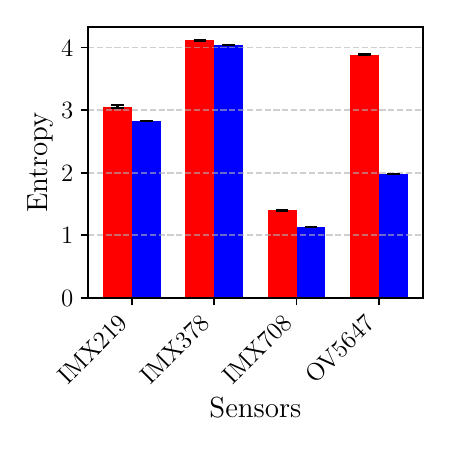}
        \caption{Entropy}
    \end{subfigure}\hfill
    \begin{subfigure}{0.24\textwidth}
        \centering
        \includegraphics[width=\textwidth]{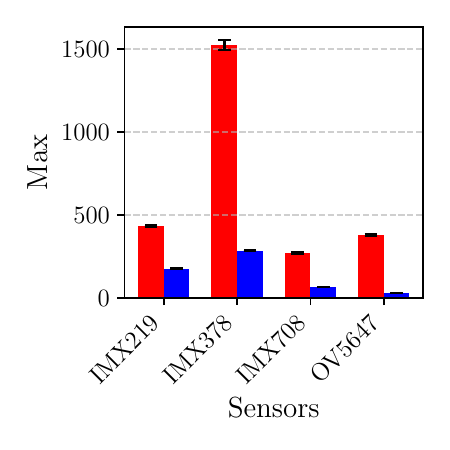}
        \caption{Max}
    \end{subfigure}
    \caption{(a) and (b): Temperature; and and (c) and (d): Sensor Types.}
    \label{fig:entr-max_temp_sensor}
\end{figure}

\end{document}